\newcommand{\CLASSINPUTbaselinestretch}{1.2}
\documentclass[technote,onecolumn,12pt]{IEEEtran}
\usepackage{algorithm2e}
\usepackage{algorithmic}
\usepackage{epsf}
\usepackage{mathrsfs}
\usepackage{subfigure}
\usepackage{graphics}
\usepackage{graphicx}
\usepackage{epstopdf}

\begin{document}

\title{
NCRAWL: Network Coding for Rate Adaptive Wireless Links
}
\author{  
Ioannis Broustis$^*$, 
Georgios Paschos$^*$, 
Dimitris Syrivelis$^*$, 	
Leonidas Georgiadis$^{\ddagger}$, 
Leandros Tassiulas$^{\dagger}$
\\
{$^*$Center for Research and Technology, Hellas \\ $^{\ddagger}$Aristotle University of Thessaloniki \\ $^{\dagger}$University of Thessaly}
 % \normalfont{Paper No. xxxxxxxx, y pages}
} 

\maketitle

%%%%%%%%%%%%%%%%%%%%%%%%%%%%%%%%%%%%%%%%%%%%%%%%%%%%%%%%%%
%%%%%%%%%%%%%%%%%%%%%%%%%%%%%%%%%%%%%%%%%%%%%%%%%%%%%%%%%%
%%%%%%%%%%%%%%%%%%%%%%%%%%%%%%%%%%%%%%%%%%%%%%%%%%%%%%%%%%
\begin{abstract} 
Intersession network coding (NC) can provide significant performance benefits via  mixing packets at wireless routers; these benefits are especially pronounced when NC is  applied in conjunction with intelligent link scheduling. 
NC however imposes certain processing operations, such as encoding, decoding, copying and storage. 
When not utilized carefully, all these operations can induce tremendous processing overheads in practical, wireless, multi-rate settings. 
Our measurements with prior NC implementations suggest that such processing operations severely degrade the router throughput, especially at high bit rates. 
Motivated by this, we design {\bf NCRAWL}, a Network Coding framework for Rate Adaptive Wireless Links. 
The design of NCRAWL facilitates low overhead NC functionalities, thereby effectively approaching the theoretically expected capacity benefits of joint NC and scheduling. 
We implement and evaluate NCRAWL on a wireless testbed. 
Our experiments demonstrate that NCRAWL meets the theoretical
predicted throughput gain while requiring much less CPU processing, compared to related frameworks. 
\end{abstract}

%%%%%%%%%%%%%%%%%%%%%%%%%%%%%%%%%%%%%%%%%%%%%%%%%%%%%%%%%%
%%%%%%%%%%%%%%%%%%%%%%%%%%%%%%%%%%%%%%%%%%%%%%%%%%%%%%%%%%
%%%%%%%%%%%%%%%%%%%%%%%%%%%%%%%%%%%%%%%%%%%%%%%%%%%%%%%%%%
\section{Introduction}
\label{sec:introduction}

Intersession Network Coding (NC) enables the local processing and mixing of independent traffic flows. 
 Combining such flows %It is possible to combine flows belonging to different sessions at different 
at wireless routers  can increase the available capacity \cite{ahlswede, proutiere, cope}.   
However, 
%all these advantages are 
 such increase is 
evident only when:   
(a) routers (which perform the encoding operations) are able to quickly identify efficient coding opportunities that increase the NC gain;   
(b) packet decoders are able to correctly decipher the encoded packets and acknowledge the decoded packets that they receive in diverse channel conditions;  
 and 
(c) the overheads imposed due to the inclusion of additional packet headers %\cite{cope} 
as well as packet processing operations \cite{memorycopy} are kept minimal. 
%Note also that while 
 While NC can increase the router throughput in random-access networks \cite{ieee80211},  
prior studies have shown that 
  when the packets are scheduled carelessly, the benefit is reduced \cite{proutiere, nococo}. 
  With multi-rate links, and when % are considered and 
  decisions are made based on statistical information,  
  scheduling is necessary to avoid 
  packet %considerable 
  losses. 
All these factors should be taken into account when designing and developing practical 
NC algorithms and systems. 

{\bf Prior  NC systems % implementation efforts 
 do not consider such effects:} %are overhead intensive:} 
Although intersession NC can theoretically offer unprecedented wireless router capacity benefits, realizing these benefits in practical systems is by no means an easy task.
Prior implemented efforts %In the first approach 
towards practical NC \cite{cope,er,clone,more}, have demonstrated %the authors were able to show 
 throughput benefits at low transmission rates but have also 
discovered a series of complexity issues arising in such implementations. 
 Our measurements across different  testbeds suggest that 
 it is very difficult for such NC %experimental 
 implementations %for NC %exists to-date that delivers 
 %are unable 
 to deliver the expected throughput gains %from NC 
  in practical multi-rate %wireless 
 deployments. % using high channel rates. 
 This is due to 
  two reasons, which motivate our study. We explain them below. 

{\bf \em a. Overhead intensive design:} 
  With NC,  routers 
 need to be %come 
 aware of the packets that have been successfully overheard by each neighbor, %; % copies that each neighbor maintains, towards making efficient encoding decisions. 
% with this,   routers %need such information, in order to 
%can efficiently 
in order to 
decide %on 
which packets to encode together %\cite{cope} 
and when.   
A method that has been commonly adopted %that has been adopted 
for addressing this requirement, % these issues 
 is by enforcing every neighbor into  explicitly acknowledging  % (\emph{explicit} ACKs) 
   overheard packets. 
 However, in certain practical settings this approach may perform poorly. 
This is because: %due to two main reasons: 
 (a) 
the timeliness of information can be severely impaired, as % for some terminals since 
it depends on random medium access; %, which is random; %The most important flaw, however, is that 
 and 
 (b) 
 additional packet processing needs to be performed,  
  %(storage, retrieval, header handling), 
 which intrusively increases the already imposed processing overhead by NC. % due to the processing operations that NC requires. 
 Due to such operations, routers become overloaded. 
 Thus, %As a consequence, 
 although the channel may be conducive to the use of high bit rates, routers may be incapable of transmitting as many packets in order to meet those rates. 
Our experiments across two different 
testbeds%\footnote{We performed these experiments using the publicly available implementation of these frameworks.} 
 (with various approaches such as \cite{cope} and  \cite{er} running on 1 GHz CPU, 1 GB RAM devices \cite{orbit}) suggest that in the simple scenario where Alice and Bob exchange packets through an 
intermediate relay  %\cite{cope} 
with all links having a PDR (Packet Delivery Ratio) close to 1, the observed CPU utilization was at 100\% in 802.11g when bit rates of 36 Mbps or above were used.  
Moreover, the maximum %achieved 
router throughput for Alice's and Bob's flows  
was apprx. 6.5 Mbps on average, compared to apprx. 8 Mbps of the pure IEEE 802.11 protocol \cite{ieee80211}. 
 These %preliminary 
 measurements suggest that  
  with such %prior implementations, 
 %when such 
 design choices, % are followed, 
  the benefits due to NC 
cannot outweigh the performance degradation due to %the 
excessive imposed overheads.  % that these choices impose. 

{\bf \em b. Absence of scheduling techniques:} 
 Previous studies have demonstrated the benefits of jointly applying NC and link scheduling  \cite{proutiere, nococo}. %however,  
 Prior practical implementations  have not incorporated any such techniques, 
 while they have not been designed to 
 %and (b) are not modular and thus they cannot naturally 
 host scheduling ideas. 
This necessitates the design of a broad, although lightweight framework, which can facilitate the efficient coexistence of NC and scheduling. 

{\bf Our contributions:} 
 We present the design and implementation of  NCRAWL, our Network Coding framework for Rate Adaptive Wireless Links.  
  NCRAWL   %
 has been optimized at each stage of NC operations 
 It is a modular tool, which can easily host the  
 %the basis for the 
 implementation of %a plurality of 
 intersession NC schemes that are either standalone, 
or tightly integrated with scheduling algorithms. %, as we explain later. 
More specifically, 
%the inherent properties of our framework, which distinguish it from other related systems are the following: 
 our framework differs from other related systems in the following aspects:
\begin{itemize}
\item {\bf  \em NCRAWL is modular:}
 %Potential NC and scheduling
 Algorithms can be easily developed as extensions to        
  NCRAWL modules.   
 These modules manage all the  NC operations, such as encoding, decoding, storage and routing %. 
 %They have been designed and implemented 
 in a lightweight manner, which allows for overhead-limited %NC and scheduling 
 network operations. 
\item {\bf  \em NCRAWL realizes joint NC and scheduling:} 
Our framework is the first to facilitate the practical coexistence of NC and scheduling. 
Theoretically shown throughput benefits can be easily assessed on NCRAWL and adapted for operating on real networks with limited effort. 
\item {\bf  \em NCRAWL uses solely stochastic information for overhearing:} 
%We do not adopt the approach of acknowledging every overheard packet to routers. 
With this we %This design choice 
avoid many %overcomes the need for 
overhead intensive %packet 
processing operations. 
%By allowing the 
The practical integration of NC with scheduling %algorithms %we combat the uncertainty and 
 provides a well performing 
 lightweight solution. 
%It also allows the  integration of NC with optimal algorithms, thus providing the best solution within the class of implicit ACK schemes, as we show in section \ref{sec:case}. 
\item {\bf  \em NCRAWL is channel aware:} 
%With our framework, w
 Routers are aware of all the potential NC opportunities that can take place within their neighborhoods at all times, as well as 
 %the maximum-throughput bit rate that can be used for transmitting encoded packets destined to neighbor groups.
 the maximum transmission rate of the encoded packets that allows for decoding.
Routers can also quickly determine which packets should be encoded together to offer the highest performance benefit. 
 %in order to obey the potential coding policy that is followed at each time by every device. 
\end{itemize}
We implement NCRAWL on %the 
 {\em Click} %modular router framework, 
 as a Linux kernel module \cite{clicksite}. 
We evaluate NCRAWL on a wireless testbed  %(deployed both indoors and outdoors) 
 through measurements with various indoor and outdoor topological settings. %%% 
Our experiments demonstrate that NCRAWL  %manages to efficiently 
identifies efficient %the cases where 
NC opportunities; % will provide benefit; 
it offers significant  throughput improvements even at high bit rate regimes, where
 prior schemes are % 
 unable to operate, due to the imposed overheads.

\textbf{The scope of our work:}
 Our focus is not on  
 proposing optimal scheduling policies for % that fully exploit %towards the full 
%exploitation of 
NC, but on developing an accurate, lightweight and easy-to-deploy system that can host NC and/or scheduling schemes. %
NCRAWL can be applied on  % WiFi 
 routers, %and 
 keeping their functionality simple and fast, %considering 
 given that they need to process and route many thousands of packets per second. 
We provide a practical and realistic design, by considering cases where encoded packets are decoded at the next hop.  

The rest of the paper is structured as follows.
In section \ref{sec:background} we discuss relevant previous studies. 
% on NC and scheduling.
%and in 
 In section  \ref{sec:nc_scheme} we provide a high level overview of the considered 
 NC scheme.
In section \ref{sec:arch} we present the modular design and implementation of NCRAWL.
%In section \ref{sec:implementation}, we describe the implementation of our framework, and we discuss its differences from COPE \cite{cope}. 
In section \ref{sec:case} we demonstrate the strengths of our %architectural 
 design through a scheduling-driven case study. 
%We also explain how such ideas can be implemented  using NCRAWL. 
In section \ref{sec:measurements}  we assess the performance of  NCRAWL  
  %in terms of achievable throughput and resource utilization, 
  via extensive measurements. %
 Our conclusions form  
 section \ref{sec:conclusions}. 
%%%%%%%%%%%%%%%%%%%%%%%%%%%%%%%%%%%%%%%%%%%%%%%%%%%%%%%%%%
%%%%%%%%%%%%%%%%%%%%%%%%%%%%%%%%%%%%%%%%%%%%%%%%%%%%%%%%%%
%%%%%%%%%%%%%%%%%%%%%%%%%%%%%%%%%%%%%%%%%%%%%%%%%%%%%%%%%%
\section{Related work}
\label{sec:background}
\setcounter{paragraph}{0}
In this section we discuss previous related NC studies and differentiate our work. 

{\bf Experimental studies on wireless coding:} 
Katti et al. \cite{cope} propose COPE, the seminal implementation of wireless NC. % network coding. 
With COPE, 
 %Their design mandates that 
  routers are fully aware of packets that have been overheard %in the past 
  by every neighbor. 
For this, each node is required to inform the router about overheard %the 
packets. % that it has overheard and stored. 
Experiments with COPE on a wireless testbed show that even with very simple encoding operations, intersession NC %network coding 
can provide significant capacity gains. 
They also study the interactions of network coding with the routing and the higher layer protocols. 
We provide more details about COPE later in this paper, where we differentiate our design and implementation. % from the one in \cite{cope}. 
Rozner et al. in \cite{er} present ER, a scheme that adopts the design of COPE and employs NC to perform efficient packet retransmissions. 
With ER, packets that need to be retransmitted are coded together, such that one retransmission can recover multiple packet losses. The authors show that the problem of selecting the optimal set of packets to code together is NP-hard; 
they propose a set of heuristics that can be followed to make coding decisions. 
Kim et al. \cite{tskim} extend the design of COPE to include NC-aware bit rate control and clever selection of nodes that acknowledge the reception of encoded packets. 
Rayanchu et al. \cite{clone} propose CLONE, a suit of algorithms for NC that take into account    losses on wireless links. However,  \cite{er}, \cite{tskim} and \cite{clone} all follow COPE's logic regarding the dissemination of information about which packets have been stored as keys. Moreover, these studies do not make online decisions about whether to enable coding or not, based on the link quality. 
MORE \cite{more} is a routing protocol %for static mesh networks, which 
where routers perform random mixing of packets %, right 
before forwarding them. 
%With this, r
 Routers that overhear the same transmission may decide not to forward the same packets. 
 Sources keep sending linear combinations of a batch of packets, until receiving an ACK from the destination. %The ACK is given priority along a ``reliable" shortest path back to the source. 
MIXIT \cite{mixit} encodes symbols rather than packets. 
Similarly to MORE, batches of packets are coded together. However, since a packet is a sequence of symbols, 
 Intermediate relays use hints from the PHY layer in order to infer which symbols within a packet are %{\em clean} (i.e., 
correctly received with high probability. %) 
%and which are %{\em dirty} (
%not deciphered. 
Relays choose a vector of coefficients at random and encode packets symbol by symbol, using only the {\em clean} symbols at a certain position. 

The above experimental approaches differ from ours in that we use solely stochastic information for overhearing instead of acknowledging each particular packet. %I.e., we use only implicit ACKs. 
This allows for efficient implementation %of the framework 
and avoids %prohibits router operations on each packet which are 
computationally expensive packet processing operations. 
%In addition, this approach creates a set of new problems regarding joint NC and scheduling with feedback. 
Our work offers a valuable %measurement 
tool for studying problems regarding joint NC and scheduling with feedback, and 
 %algorithms in this direction. 
  can potentially be intertwined with an optimal algorithm to provide the best solution within the class of implicit ACK schemes. 
  %
  %
  %
%  In this paper, we
 We showcase the operation of NCRAWL with various heuristic lightweight  algorithms. % which perform well. 
As we show  in  section \ref{sec:measurements}, there are cases of implicit ACK schemes where NC incurs throughput loss, unless careful scheduling is used.   Making decisions based on multiple rates is another important innovation of our framework. 
 To the best of our knowledge, our work is the first to provide a coherent, lightweight framework for practically assessing joint NC and scheduling schemes. 

{\bf Analytical and simulation NC studies:} 
Chaporkar and Proutiere \cite{proutiere} show that systems with NC may actually have smaller throughputs than if coding is not applied. 
They show that unless appropriate scheduling is applied, NC may lead to performance degradation. % in many topological scenarios. 
 We support this claim by identifying an additional example when implicit ACKs are used. In the same work, 
 Chaporkar and Proutiere propose a generic framework that characterizes the throughput region with NC and enables the design of adaptive joint NC and scheduling schemes. 
Finally, 
they propose XOR-Sym, a computationally simple 
NC scheme that can be applied to symmetric routes. 
With XOR-Sym, packets are decoded at their destinations only and not at intermediate nodes along a path. This protocol %has similar differences to our work as COPE has and it 
considers only symmetric flows disregarding opportunistic listening. 
On the contrary we focus on exploiting opportunistic listening. 
Liu and Xue in \cite{chliu} consider NC for two-way relaying in a three-node network. They analytically characterize the achievable rate regions for the traditional Alice-Relay-Bob topology, and they find the theoretically optimal end-to-end sum rates. 
Scheuermann et al. \cite{nococo} propose noCoCo, a deterministic packet scheduling scheme for NC within two-way multihop traffic flows. 
Their scheme involves per-hop packet scheduling, NC and congestion control. 
%Seferoglu et al. \cite{athina_video} propose code selection schemes for network code selection that consider the properties of video traffic. Moreover, 
Seferoglu and Markopoulou in \cite{athina_distr_video}  provide an understanding of the interplay between application data rate control and NC. 
%Furthermore,
Finally, Vieira et al. \cite{mario_globe07} %perform simulations to 
provide observations on	how the combination of NC and bit rate diversity affects the performance of practical broadcasting protocols.
They show that it is possible for multi-rate link layer broadcasts and NC to jointly increase the network throughput in multicast applications. 
%%
%Le et al. \cite{lui08} provide an upper bound on the number of packets that can be coded together, in any possible coding structure.
 %However, unlike our work, none of these studies perform real world experiments. 
More theoretical results can be found in \cite{ncpage} with the list being non-exhaustive.
%%%%%%%%%%%%%%%%%%%%%%%%%%%%%%%%%%%%%%%%%%%%%%%%%%%%%%%%%%
%%%%%%%%%%%%%%%%%%%%%%%%%%%%%%%%%%%%%%%%%%%%%%%%%%%%%%%%%%
%%%%%%%%%%%%%%%%%%%%%%%%%%%%%%%%%%%%%%%%%%%%%%%%%%%%%%%%%%
\section{Network coding scheme}
\label{sec:nc_scheme}

\begin{figure}[t]
\centering
\includegraphics[width=2in]{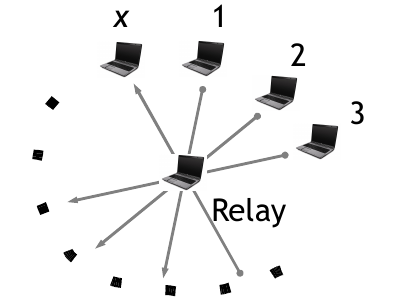}
%\label{figure:wheel}
%\vspace{-0.1in}
\caption{The $\frac{x}{2}$--wheel setting, with $x+1$ nodes and $\frac{x}{2}$ activated flows.}
\label{figure:wheel}
%\vspace{-0.12in}
\end{figure}

We study the generalized $N$--wheel topology having $N+1$ nodes as shown in figure \ref{figure:wheel}. The central node, called the \emph{relay} %(another name for it is router), 
 (or router),  
is connected to all other $N$ nodes, called \emph{neighbors}. Links between neighbors may exist as well. Each link connecting nodes $i$ and $j$ is characterized by two channel rates $r_{ij}$, $r_{ji}$ and two probabilities $q_{ij}$, $q_{ji}$ which correspond to the packet delivery ratios in each direction. 

In the above topology we focus only on 2--hop flows having neighbors as source and destination and the relay as the intermediate node. In the uplink part (the first hop of these flows), the packets are transmitted without NC towards the relay. In the downlink part, the relay selects a number of packets, applies the XOR operator and transmits the encoded packet. If a receiver recognizes its address in the header, it attempts to decode the packet in order to obtain the intended packet. To achieve this, it should have all the other packets in its buffer, in order to apply the XOR operator again. These packets are known to the receiver either because they have been generated by it (in case of symmetric flows) or they have been obtained by means of opportunistic listening, as explained above. Whenever the packet is successfully decoded, an acknowledgment message is sent back to the relay at layer-3.

Coming back to the $N$--wheel topology, in order to experiment on high gain topologies, we can impose the extra constraint that $N$ symmetric flows are defined by splitting the neighbors in two equal sets, the source and the destination set, selecting a matching of these two sets which corresponds to $\frac{N}{2}$ flows, and finally create another  $\frac{N}{2}$ flows by inverting the roles of source and destination. If only the initial $\frac{N}{2}$ flows are enabled, we refer to $\frac{x}{2}$--wheel setting else if all $N$ flows are enabled, we refer to $x$--wheel setting. Evidently, $2$--wheel corresponds to the well known Alice-Relay-Bob topology. Throughout the paper, we also use $\frac{4}{2}$--wheel referred to as half-cross and $4$--wheel referred to as cross, as well as a 
$6$--wheel where we activate the flows one after the other. It should be noted that NCRAWL supports any random subgraph of the $N$--wheel topology with any possible set of flows activated on top of it. In addition, it supports settings with any possible combination of link qualities and/or channel rates. 
%Finally, in an ad hoc setting, any node can dynamically play the role of relay. 
%
The wheel topology is the most general topology to be considered around a single node. Any actual network topology can be reduced to a wheel topology if nodes and links irrelevant to NC on a relay are removed. Since our scheme runs on all nodes in the network, this implies that our scheme works with any arbitrary network topology.
The encoding opportunities at each node are automatically discovered by the combination of NCRAWL and SRCR. Thus NCRAWL operates under any assumed graph providing opportunistically throughput gains. We study the wheel setting where the maximum such gain arises in order to showcase that NCRAWL can achieve it in many cases.

%{\bf The limitations of our scheme:}
We have incorporated the 
 %The 
 following features in order to enhance the practicality of NCRAWL. 
 As a tradeoff,  such features may limit the performance of our framework against the maximum theoretical performance of intersession coding. 

\begin{itemize}
\item The XOR operator is used instead of linear coding. It is known that the capacity region of XOR NC schemes is a subset of the one achieved with linear coding.
	\item We %avoid routing of encoded messages, by enforcing 
  enforce the decoding of encoded packets at the next hop, since this is practically the most possible case in today's wireless access deployments.  % is not allowed. Thus, for example, the butterfly topology cannot be built.
	\item Only native packets are allowed to be stored as keys. This might reduce the capacity region as well.
		\item We use implicit ACKs of overheard packets. The capacity region is reduced in comparison to explicit ACKs whenever the overhearing probabilities are small.
	
\end{itemize}
The decision on imposing these features is  justified by the necessity to keep the NC scheme simple, practical, implementable and efficient in terms of processing overhead.
%%%%%%%%%%%%%%%%%%%%%%%%%%%%%%%%%%%%%%%%%%%%%%%%%%%%%%%%%%
%%%%%%%%%%%%%%%%%%%%%%%%%%%%%%%%%%%%%%%%%%%%%%%%%%%%%%%%%%
%%%%%%%%%%%%%%%%%%%%%%%%%%%%%%%%%%%%%%%%%%%%%%%%%%%%%%%%%%
\section{Architectural Blueprint}
\label{sec:arch}

In this section, we describe the modular design and implementation of NCRAWL. 

{\bf Employing Click as the basis of our framework:} 
The main NCRAWL system has been developed in the \emph{Click} modular
router framework \cite{clicksite}. Click can be used to develop primarily OSI layer 3 
packet processors, which can be directly deployed inside the standard Linux network stack. 
A {Click} processor is mainly comprised of (a) processing stages which are called 
\emph{elements} and (b) an  % respective 
element interconnection configuration that indicates the processing flow. 
Execution in {Click} is event-driven with 4 different types of asynchronous events, namely the {\em incoming packet} event, the {\em ready-to-forward packet} event, the {\em timer expire} event and the {\em external read or write} events to {Click} memory. The first two events require  
 %the respective 
  some handling code 
to deal with network packets; %however, 
 this is not always necessary for serving the rest of the event types. 
Since all {Click} events are asynchronous, a {Click} packet processor typically % should 
 features internal queues to temporarily store %arriving 
 incoming 
 packets.
% because they are going to be forwarded asynchronously.

In what follows, %we present all aspects of the NCRAWL framework design. 
%We first discuss % starting from 
%the respective network protocol. Subsequently %We then 
 we describe the NCRAWL system design
and implementation.  %modular router 
%and finally, we
We also present the NCRAWL
interface that can be used to %(a) 
 develop new algorithms, as well as for % and (b) 
 deploying and managing %respective 
 experiments on   wireless testbeds. 

%\subsection {NCRAWL protocol}
%% Dimitri: what protocol are you referring to? 
\subsection {Design preliminaries}

\begin{figure}[t]
\centerline{ \includegraphics[width=4in]{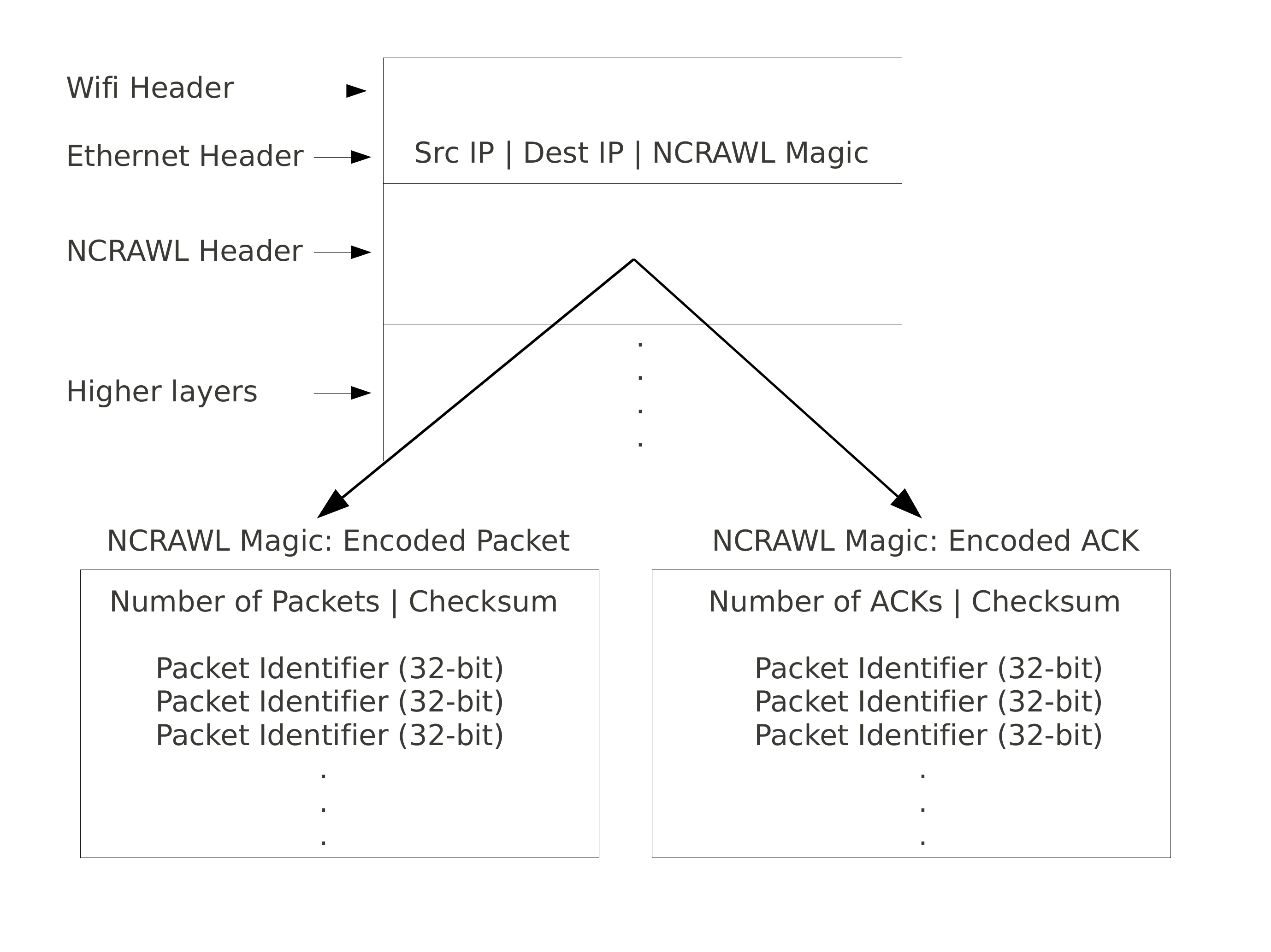}}
\caption{NCRAWL header format. The Ethernet header magic number distinguishes
between encoded packets and ACK headers which are otherwise identical.}
\label{NCRAWL:headers}
\end{figure}

NCRAWL realizes %also comprises 
an OSI layer 2.5 protocol that lies immediately under
the %adhoc mesh network 
routing layer. More specifically, it can be considered 
as an extension to the {Click} modular router implementation of the
SRCR   protocol \cite{roofnet_wiki}, 
which  is the heart of the {MIT Roofnet} %802.11  
 wireless %mesh 
 network. 
Since NCRAWL operates below the routing layer, encoded packets %cannot be
are not forwarded by a %mesh 
 node-relay; they are % and can be 
decoded at the next hop. %by immediate neighbors only. 
This
simplifies the format of the NCRAWL packet headers:  they are now used only
to encapsulate encoded packets and transfer the 
respective acknowledgments for %the 
 successful decoding %of packets  
%successful packet decode 
back to the sender. %More specifically, both 
 Both
types
of the NCRAWL header format are depicted in figure~\ref{NCRAWL:headers}.
Network wide unique 32-bit packet identifiers are made by applying the 
\emph{sdbm} \cite{sdbm} hashing algorithm on data tuples, comprised of 
packet source IP, the IP header sequence number and the respective offset. 
%\endnote{to avoid producing the same identifiers for fragmented packets%}.

\subsection {NCRAWL System Description}

%In what follows, we 
 Next, we discuss the modular design of NCRAWL in detail. 

\begin{figure}[t]
\centerline{ \includegraphics[width=3.7in]{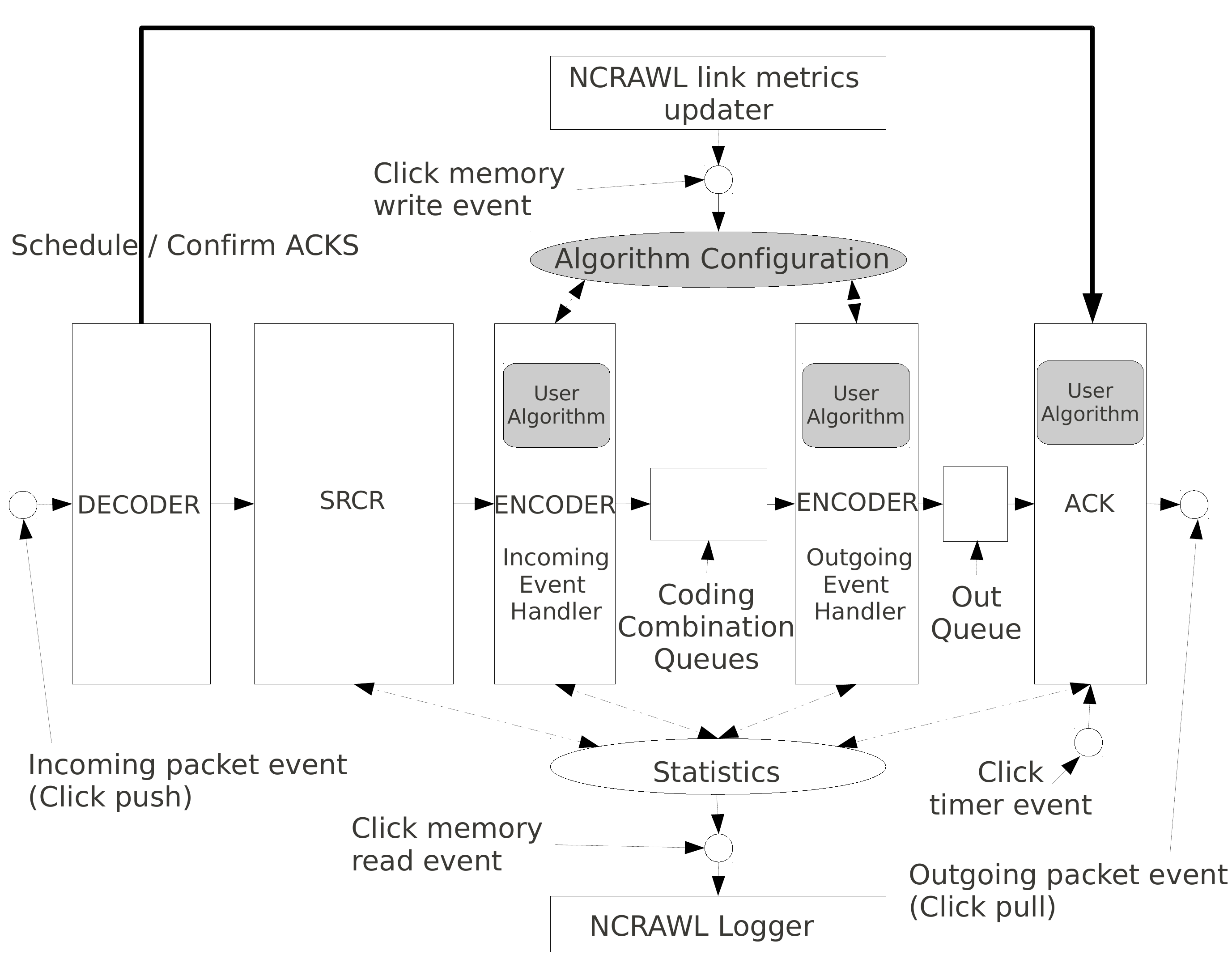}}
\caption{NCRAWL framework system components and their interaction}
\label{NCRAWL:system}
\end{figure}

{\bf The big picture:} 
The main NCRAWL system is a Click network packet processor that includes the {SRCR} %adhoc 
routing protocol implementation for wireless networks \cite{roofnet_wiki}.   We have included two additional
processing stages: % have been included, 
 %namely 
 the NCRAWL decoder and the NCRAWL encoder.
 We have developed these stages %have been developed 
 as individual {Click} {elements}, and we have placed them % and have % which have 
%been
%respectively 
%placed 
before the beginning and after the end of the {SRCR} processing
flow, respectively, as depicted in figure \ref{NCRAWL:system}. 

{\bf a. The packet decoder:} 
 The main tasks of the decoder module are the following: 
 \begin{itemize} 
\item To use the available (from overhearing or ownership) key packets in order to decode the received encoded packet. 
%\item To decode any received 
%encoded packets, as long as the proper keys are available. Such appropriate stored keys, learned through continuous overhearing, are used to apply the decoding function on the received encoded packets. 
\item To schedule the transmission of layer-3 acknowledgments (ACKs) for the correctly retrieved native packets, % decoded packets (these packets comprised the encoded packet and were retrieved 
 derived by the decoding operation.
%\item To check if expected and verify any received acknowledgements.  %and 
\item To determine any potential pending acknowledgments, as well as to verify any received acknowledgments.
\item To tag and  store all the correctly overheard data packets as potential keys; as discussed above, these will be potentially used in the near future for decoding received encoded packets. Moreover the
key repository is used for packet retransmissions,  in case an % respective 
 expected 
 acknowledgment never arrives.
\end{itemize}
The   decoder %element 
 resides at the packet receiving side of the system and %it 
 is invoked by the corresponding packet arrival event. 

{\bf b. The NC packet encoder:} 
 The NCRAWL encoder element resides at the sending side of the system and is more
complicated, since % because
 it maintains % contains 
 and manages the processor packet queues. %As a result a
 A  part of the element handles incoming packet events, another part deals with outgoing
packet events, and there is also code that gets invoked upon timer expiry  %by timer 
as well as read and write
{Click} configuration events (figure \ref{NCRAWL:system}). It is this element that exports the framework {API} 
which can be used to develop NC algorithms. % and, in 
Specifically, the main assigned tasks for this module are the following: 
%are: 
\begin{itemize}
\item 
To process and place %arriving 
incoming native packets (keys) into particular %the appropriate 
 %available 
  maintained queues. %, based on the user defined scheme. 
Our system supports a plurality of queueing operations, which can be configured 
as per the requirements of the {NC} algorithm under development.
\item
To identify and combine packets together, towards forming encoded packets. %, % as indicated by the algorithm. 
The selection of the appropriate packet set follows the directions of the NC algorithm under consideration, supported by NCRAWL. 
\item 
To piggyback any acknowledgments  
(through the use of scheduled, upcoming data packet transmissions) 
that have been scheduled by the decoder element. %,  regarding outgoing  native data packets.  
\item
To generate   potentially 
expected acknowledgment tokens for each   of the packets of an encoded combination.
\end{itemize}
%%%%%%%%%%%%%%%%%%%%%%%%%%%%%%%%%%%%%%%%%%%%%%%%%%%%%%
%

{\bf c. Maintaining up-to-date topological information:}
The \emph{link metrics updater} is responsible for collecting information about the
 current neighbors 
 %network neighborhood 
  as well as the corresponding link transmission rates and PDR values. % from all nodes. 
  This  information is gathered and passed to the rest of the system via the {Click} memory write event mechanism. Furthermore, the code that configures the encoding combination policies is invoked 
%to react 
 as needed. 

{\bf \em Gathering link quality information:} 
The {NCRAWL} updater relies on the existing {SRCR} protocol component, which maintains 
 %the neighborhood view 
 link connectivity information and performs periodic measurements on all links. {SRCR} sends probe packets at all rates to determine the PDR for each link and chooses the highest rate that performs well. PDR information is then used by {SRCR} to calculate the {ETX}  or ETT metric \cite{etx, ett}, which %encapsulates 
 provides information about entire routing paths (not just 1-hop links).  
  This information is kept in the {SRCR} link table, and is accessible by our %other 
   {Click} components. The {SRCR} measurement period can be set as desired  (the default value, also used in our work, is 3 sec). 

{\bf \em Managing neighbor information:} 
Based on the information stored in the {SRCR} link table, the %{NCRAWL} 
 link updater maintains %in memory 
  its own so-called {Neighbor Table} (NT), which includes information for its %(direct) 
  neighbors. Initially, the NT is empty. The updater periodically reads the {SRCR} link table 
  %(using the same period as {SRCR}) 
  and updates NT as needed. %Each time the 
   The  NT contents are updated %change %, i.e., 
   whenever (i) a new neighbor appears,   (ii) an existing neighbor disappears, or (iii) a certain link quality changes. 
   In such cases, the {NCRAWL} updater broadcasts a packet with the new NT contents and sets a timer. When such a NT packet is received (overheard), the updater replies by broadcasting its NT, provided it has not done so recently. The reply suppression threshold is set equal to the {SRCR} period.
The NT packets are used by the {NCRAWL} updater to maintain the so-called {Received NT Table} (RNTT). This table complements the NT, holding information about the link quality as  
 %experienced/
   measured by the neighbor nodes \emph{themselves}. 
 % rather than by the local node. 
  When an NT packet arrives, the corresponding RNTT entry is updated.  % respectively. 
   Packets from nodes that are not in the NT are ignored; a node must be ``officially'' reported as a direct neighbor by {SRCR} in order to be considered by {NCRAWL}.

{\bf \em Feeding NC algorithms with updated topological information:}  
Each time the updater modifies the contents of the NT or RNTT (i.e., each time it proactively sends   
 %a NT packet 
 or receives a NT packet which leads to an update of the RNTT) a timer is set. 
  %When the timer expires, 
  Upon timer expiry,   
  the new link qualities are passed to the main {NCRAWL} system,  
where they will potentially drive adaptive NC decisions, based on the NC designer's needs. 
This timeout is (generously) set to 1 sec, providing ample time for any NT reply packets to arrive. 

{\bf \em Keeping overheads low:} 
The NCRAWL updater employs its own threads of execution to perform these information maintenance tasks (the current implementation uses 2 threads), but these remain suspended most of the time, making this component quite unintrusive in terms of CPU occupancy. Moreover, only a small fraction of the wireless bandwidth is typically used to collect the required link quality information from the neighboring nodes. Finally, the ``reactiveness'' of the updater is a function of the {SRCR} period. If a smaller period is used, link changes can be tracked faster (and more accurately) but the processing and communication overhead will increase too. 
Note that the implementation of the {NCRAWL} updater 
  %is not optimized for 
  %rapidly changing environments as this is not the focus of this work.
  can be trivially adjusted to cooperate with other link information gathering protocols as well,  
  i.e., it is not tied to SRCR. 

%%%%%%%%%%%%%%%%%%%%%%%%%%%%%%%%%%%%%%%%%%%%%%%%%%%%%%
%
{\bf d. NCRAWL logger:}  
 The  {\em read} events are used by another application, the %so-called 
NCRAWL {\em logger}, which gathers
various statistics that are generated online by both the encoder and decoder.  % elements. 

{\bf e. NCRAWL acknowledgments:} 
 %Regarding the acknowledgement scheme,  
 NCRAWL acknowledges individual (native) data
packets, but also groups packet acknowledgments per encoded combination. 
 With this, if the same encoded packet has been successfully decoded at one recipient but failed at
another, the sender can figure out which of the undelivered packets can be reused in encoded
combinations, based on whether they have been logged successfully as keys by fellow recipient
nodes. %We should note here 
  Note that NCRAWL provides this support; however,  
%described information and
 it expects that the user algorithm will make the final scheduling decisions. 
NCRAWL uses by default a user defined timeout threshold to resend 
packets that have not been acknowledged. Note also that 
 %Finally,
a timer-expire event triggers the transmission of acknowledgments in separate packets when there is not
enough outgoing traffic to piggyback them (figure \ref{NCRAWL:system}). NCRAWL also % with the required pace and also
reschedules packets from the key repository for which acknowledgments have not arrived.

{\bf \em Utilizing resources effectively:} 
 Efficient resource utilization 
 %was a main concern during the design of the various NCRAWL subsystems. 
 is an inherent property of NCRAWL.  
 %More specifically, the 
  The repository that stores copies of packets
uses a FIFO queue as the main indexing mechanism and can host up to a user defined quantity.
After the storage limit is reached, the oldest packet is removed in order for a new one
to get stored. The same packets are also indexed in a hash table based on their network-wide
unique identifiers, as % that 
  we  %have 
  previously discussed. The hash table is used to quickly
retrieve packets either as keys for decoding, or for resending them in case an expected
ACK token expires. The same indexing approach has been used for the
ACKs and expected ACK tokens as well.

\subsection{Implementing {NC} algorithms} 

  NCRAWL exports an API (Application Programming Interface) that can be used to implement scheduling algorithms for intersession NC.
  This API is %comprises 
  a library of functions that can
be used to carry out NCRAWL
common tasks and mandatory function extensions to the handling code of each event. Points of extensibility and/or programmability are denoted in figure \ref{NCRAWL:system} with shadowed   boxes.

{\bf Implementing packet handling operations:} 
% For 
 Regarding 
the incoming packet event, %extension 
 the designer should account for placing arriving packets 
into  proper queues. 
In particular, 
 each flow is associated with a queue and the scheduler checks all available controls, i.e. activates  a set of queues by combining one packet from each queue. It maintains a list with the expected score (reward) of all controls and selects one of those controls at each time instance. The controls that activate only one queue correspond to the case of transmitting non coded packets. It is always possible to deactivate NC by imposing the use of only those latter controls. 
 With this,  
 the developer may implement logic that disables NCRAWL 
when needed. Furthermore, after the placement of the incoming packet, the developer may: 
 (a) invoke the function that chooses the next encoded packet queue combination according 
to the NC algorithm, 
 (b) retrieve the packets from the respective queues, 
 (c) encode them and schedule the encoded packet for transmission by placing it into the outgoing queue. 
 This operation may be repeated until the outgoing queue contains a user
defined number of packets. Apart from the queue combination retrieval 
function, which needs to be implemented by the algorithm developer, the rest 
of the required functionality is already seamlessly provided by NCRAWL.

{\bf Implementing the core NC logic:} 
The main algorithm implementation takes place in the context of the {Click} memory 
write event, generated by the NCRAWL updater. The latter provides 
the user with a table of single direction links with entries denoted by the corresponding  
source and destination IP combinations. Each entry holds the link direction 
{PDR} and transmission rate. 
The provided information can be used by the network 
 coding algorithm to decide the valid NC combinations  
 by selecting the queues to activate together.
 %by grouping  the respective queues. 
%
 Since this part of the code runs periodically, the developer 
is encouraged to implement any complex algorithm steps here and thoroughly
index the NC available combinations.  With such an approach, the overhead of choosing 
the most beneficial packet combination during the incoming or outgoing packet events 
will be minimized.

{\bf Sending data and ACK packets:} 
The outgoing packet event checks the size %number of packets that wait on 
 of the outgoing packet queue. 
 If  this is %they are 
 below the   defined threshold, the functions that choose and encode 
combinations are called 
 in the exact say way as for 
 %exactly as it happens during 
  the   incoming packet event. Then the
next packet on the outgoing queue gets transmitted. 
The developer may also add logic for the handling of ACKs. By default, 
NCRAWL resends packets that have not been acknowledged, by directly placing 
them on the outgoing queue. 
 It is possible, however, that an algorithm deals with the failed packets (see section 5 for an example). 
 Since the ACK scheme groups ACKs 
that belong to the same encoded packet, the developer knows which packets have been decoded
successfully at which destination, and may extend the NC algorithm to decide
whether a packet should be resent directly, or reconsidered for encoding combinations.

{\bf System monitoring:} 
 %Furthermore, 
 %NCRAWL 
 Our framework  
 allows %also exports support for 
  for 
 user defined timer events. 
% Regarding the statistics, the framework logs with counters all the incoming and outgoing packet activity and lengths of all queues.
Statistics for incoming--outgoing packet activity as well as queue lengths are all logged using counters.
The NCRAWL logger periodically retrieves %the 
statistics %information 
and 
notifies the user at runtime about the flow stability and the corresponding queue lengths. The latter are 
also available for use in the NC algorithm if needed. Finally, the developer may also 
implement additional debugging support for inspecting the algorithm configuration at runtime, using 
the standard Click support for the read handler. 

\subsection{Deploying NCRAWL Experiments}
\label{subsec:deploy_NCRAWL_experiments}

We have integrated NCRAWL deployment scripts with the OMF framework \cite{orbit} for wireless testbeds. {OMF} is a Control, Management and Measurement Framework that provides %the  
 users with   tools to describe, execute and collect experimental results %of an experiment 
 in a straightforward manner. There are three main components that comprise OMF; we   describe them below: %(a) the gridservices, (b) the nodehandler and (c) the nodeagent. Below, we provide a short description for each one of these components.
\\
$\bullet$ {\bf \em Gridservices:} This is a set of web services that are used by {OMF} to fetch information and perform actions remotely on the nodes. These services can be used for loading the system image to nodes, executing   experiments and   collecting   results.
\\
$\bullet$ {\bf \em Nodehandler:} This component resides on the central server that interacts with the user for the experiment submission. Moreover, it provides the necessary applications for node system image loading, experiment execution, image saving and node status check. This component communicates with both the gridservices and the nodeagent to get the required information and perform actions. Regarding the experiment deployment, nodehandler contains a set of prototypes that can be used for experiment definition. It also %Then it observes the experiment execution and 
notifies the user for any problems that may arise.
\\
$\bullet$ {\bf \em Nodeagent:} %A nodeagent instance is deployed on each testbed node. 
%Contrary to the nodehandler which is triggered only upon a load, execution, save or status call, the nodeagent (deployed on each testbed node) is constantly active. Its 
The task of this constantly active component is to wait for information  that contains instructions for the experiment deployment,  arriving  from the nodehandler.

 NCRAWL extensions have been written for both nodehandler and nodeagent. The former   
performs transfers of the Click executable along with user defined parameters. 
%
% Yannis: what does the following mean?
%
   %like the 802.11 channel  that should be used.
% and others.  Ti simainei to and others???????????????????????
 The latter retrieves local node information (e.g. the network interface name and MAC address)  and then   parametrizes a generic NCRAWL deployment script that gets immediately invoked to start the local Click NCRAWL instance. 
Finally, the nodehandler is notified if the deployment was successful; Upon success, %in such a  case 
 experiments can start. We have written nodeagent scripts to deploy \emph{iperf} instances and collect results at the nodehandler. 
With %the 
OMF, % management system, 
NCRAWL experiments can be deployed %in two simple steps 
 with minimal user effort. %interaction.

%%%%%%%%%%%%%%%%%%%%%%%%%%%%%%%%%%%%%%%%%%%%%%%%%%%%%%%%%%
%%%%%%%%%%%%%%%%%%%%%%%%%%%%%%%%%%%%%%%%%%%%%%%%%%%%%%%%%%
%%%%%%%%%%%%%%%%%%%%%%%%%%%%%%%%%%%%%%%%%%%%%%%%%%%%%%%%%%
\section{Case study}
\label{sec:case}

As discussed above, the relay maintains a queue for each pair of source-destination in the neighborhood that lacks direct connection (2--hop flows traversing the relay). At each transmission slot, the scheduler should select a number of packets, encode them and send the encoded packet to the MAC layer for transmission. The problem is then of scheduling nature; to select a number of packets for transmission. Note that selecting only one packet corresponds to transmission without NC. This section demonstrates by example how the NCRAWL framework allows for easy implementation of scheduling algorithms and presents a case study to be used for the performance comparison of the next section.

A well known family of optimal algorithms for scheduling is the maximum weighted matching algorithms, applied in the stability theory of stochastic networks and input queues switches, see for example \cite{TW92,MMAW99}. 
In these algorithms, the available control actions are chosen to maximize a reward which depends on link rates and queue lengths. 
The application of such algorithms for the solution of the joint NC and scheduling problem with arbitrary rates is then promising (see \cite{Eryilmaz07controlfor} and papers in which it appears as a reference).

For the case of the implicit ACKing scheme, used by our framework, one observes that the rate of service for packets of a particular source-destination pair is actually random. The randomness comes from the fact that some packets needed for decoding may be missing because of an overhearing failure. Nevertheless, a NCRAWL equipped relay owns the probabilistic information of overhearing and is then in position to determine the probability of decoding and thus the average service rate. A control action consists of selecting a number of queues to serve at a single decision instance. The reward of each control is the sum of queue length times the average service rate for each queue activated by the control. The average service rate is the expected number of successfully serviced packets times the transmission rate for which the encoded packet can be received by all intended receivers. This rate is the minimum of the reception rates of all intended receivers. We write: 
%
%\vspace{-0.1in}
\begin{equation}
%\vspace{-0.1in}
\mu_i(C)=\prod_{j\in C}q_{s_j,d_i}\min\{\mathbf{r}_C\},
\label{equation:service_rate}
\end{equation}
where $C$ is the control, $q_{s_j,d_i}$ is the probability that the destination of flow $i$ overhears the source of flow $j$ at the uplink phase, $\min\{\mathbf{r}_C\}$ is the transmission rate of the control and $\mu_i(C)$ is the expected service rate of flow $i$ when the control $C$ is selected. Here we have assumed that the overhearing events are independent. This is a realistic assumption since the time and space for each overhearing event is different. Collisions and Rayleigh fading may be the causes for this randomness.

\subsection{Algorithm 1}

Let $Q_i$ be the queue backlog at the relay for the flow $i$. We are then in position to design our first algorithm.
%\vspace{-0.15in}
\begin{algorithm}[ht]
\small{ 
%\SetLine
\KwIn
{$Q_i, \mu_i(C)$}
\KwOut
{$C^*$}

$w_{\max}:=0$\;

\For
{$C\in \mathcal{C}$}{

  $w(C):=\sum_{i\in C} Q_i \mu_i(C)$\;
  
	\If
	{$w(C)>w_{\max}$}{ $C^*$ :=
    $C$\; } 

}
\caption{MaxWeight Algorithm without feedback}
\label{alg:1}
} 
\end{algorithm}
%\vspace{-0.15in}

An issue raised in this algorithm is the fact that the number of possible encoding combinations to be examined is exponential in nature. If, for example, we assume that overhearing is possible for all receivers except the destinations, then the number of combinations is actually $2^N-1$ where $N$ is the number of source-destination pairs (or 2--hop flows). The question is then whether the computational overhead for the weights is prohibitively high. In subsection \ref{sec:case-implement}, we explain how the list of weights is maintained in order to reduce the number of calculations per slot.

The algorithm 1 is throughput optimal under the condition that the knowledge of the aforementioned probabilities of overhearing cannot be altered during transmissions. This happens when (i) the probabilities are 0 or 1, as in Alice-Relay-Bob topology (and any other symmetric flow setting) or (ii) upon a decoding failure we reschedule the uplink transmission for the failed flow. The latter may arise in a TCP scenario. In the general case, however, whenever a particular encoded packet is not correctly decoded, the packet remains in the queue at the relay but extra feedback information is obtained. If for example $P_1 \oplus P_2$ is not decoded by both receivers, the relay knows that these two packets are not overheard by receiver 2 and 1 respectively, and the proper action is to correct the overhearing probabilities to zero and never encode these two specific packets again. The impact of feedback clearly biases the probabilities of decoding.
 The knowledge state of each packet evolves in a such a way that future states depend on the control action selected at present and %, %a problem that is not dealt in the framework of stability theory in stochastic networks. As
 as a result, in the general case, not only algorithm 1 is not optimal but it might perform quite badly when the overhearing probabilities are quite small.

\subsection{Algorithm 2: the case of two queues}

Another idea is to propose an algorithm which is not necessarily optimal, but manages to handle the feedback information successfully. In general, an algorithm should be able to predict the future effects of current control actions. Here we restrict our search to the category of the so-called myopic algorithms, trying to solve the problem given only the current state and disregarding the future. We  consider the %simple 
problem of mixing only two flows. 

In order to cope with feedback, we add two more knowledge states. Apart from newly arrived (unknown) packets whose behavior is captured by known probabilities, we have a state for ``good'' packets (overheard by the other receiver) and one for ``bad'' packets (those not overheard by the other receiver). Thus, the system maintains the queues $Q_i^s$ where $i\in\{1,2\}$ signifies the flow and $s\in\{u,g,b\}$ signifies the state. The set of controls $\mathcal{C}$ contains all controls that activate one or two queues with the constraint that no two queues from the same flow can be activated.
%\vspace{-0.15in}
\begin{algorithm}[h]
\small{ 
%\SetLine
\KwIn
{$Q_i^s, \mu_i^s(C)$}
\KwOut
{$C^*$}

At feedback time:
	\begin{itemize}
		\item For each packet that was not correctly decoded define whether it is good or bad.
		\item Bad packets are directly sent to the MAC layer for transmission without coding.
		\item Good packets are sent to the corresponding queue at the good state.
	\end{itemize}
At decision time:

$w_{\max}:=0$\;

\For
{$C\in \mathcal{C}$}{

  $w(C):=\sum_{C} Q_i^s \mu_i^s(C)$\;
  
	\If
	{$w(C)>w_{\max}$}{ $C^*$ :=
    $C$\; } 

}
\caption{Myopic Algorithm with feedback}
\label{alg:2}
}
\end{algorithm}
%\vspace{-0.15in}
The packets are initially injected in the queues at the unknown state. Once a packet is not decoded properly, the relay classifies it as either good or bad based on feedback information. If bad, it is retransmitted without encoding (thus the queues $Q_i^{b}$ are not needed actually). If it is deemed as good, it is transferred to the corresponding queue at the good state ($Q_1^{g}$ or $Q_2^{g}$ depending on the flow it belongs to). When calculating average service rates, the packets at the good state have probability of overhearing equal to one. Apart from these alterations, algorithm 2 works in the same way as algorithm 1.

In \cite{rawnet} it is shown that an enhanced queue length based algorithm solves optimally the joint NC and scheduling problem arising in intersession coding at the relay node. This solution might be costly in terms of resources, and therefore suboptimal algorithms might be preferred. For this reason, our framework serves as an ideal substrate for performing measurements of such algorithms.

\subsection{Algorithm 3: fixed threshold policy}

For reasons of performance comparison we define a third algorithm. This algorithm operates only with implicit ACKs and makes decisions based on principles used in the COPE framework. In this sense, it emulates COPE in its probabilistic mode. The important differences to our algorithm are that instead of calculating average service rates, the $\delta$--Fixed Threshold Policy ($\delta$--FTP) simply marks the incoming packets with information about decoding opportunities. In order to do so, overhearing probabilities $q_{i,j}$ are compared with a fixed threshold $\delta\in[0,1]$ and set to $1$ if they exceed the threshold or zero otherwise. 
The algorithm selects at each decision instance the control that maximizes the number of transmitted packets.

\subsection{ NCRAWL algorithm implementation}
\label{sec:case-implement}
Next, we demonstrate how to implement the three above algorithms on NCRAWL. 
For all three cases, 
 %In all aforementioned algorithm cases 
 we configure NCRAWL at each node to maintain  
 %(in the context of each relay) 
 one queue per flow for incoming packets.%, per flow. 
 %This is where incoming packets get stored. 

{\bf Implementing algorithm 1:} We first describe how one may organize queues in an efficient manner. Subsequently, we show how to utilize the queue information to apply NC. 
\\
$\bullet$ {\bf \em Organizing packet queues:} 
To begin with, %we store each feasible 
%combination of flow queues into a separate vector. 
we dedicate one vector per control which contains % In other words, every such individual vector contains 
the identity of the involved queues (e.g. the flow it belongs to and/or the state) and the identities of the packets enqueued at the involved queues. 
 %a data tuple. 
The formed vectors are stored in % tuples are grouped together on 
 a double linked list. 
 Each vector is assigned a weight (or reward); the higher this weight, the higher the   
 preference of the encoder for using the combination.  
This weight is recalculated every time the backlog size of 
 a %respective 
 member queue %backlog 
 changes. 
    The linked list is formed such that the head of the list contains always the 
 current maximum
 weight. %, with subsequent list members following in a decreasing weight order.   % combination which will be used for the next packet transmission. 
 For the sake of  low processing overhead, 
%Moreover, the 
 vectors % combination data tuples 
 are also directly indexed by their member %flow 
 queues; with this, % so that 
the weight update process is fast. 
 As one may expect,  vectors as well as their linked list 
%
%The described data structures 
 are all constructed during the NCRAWL updater write event. 
\\
$\bullet$ {\bf \em Applying NC operations:}  
 Given the construction of the control list, 
the encoder event examines the head of the list, and further: 
(a) retrieves packets from their respective queues, 
 (b) updates the %combination 
 vector weights (since the respective backlogs are decremented), 
and 
(c)  sets the vector with the highest weighted combination as the head of the list. 
The latter is actually
a process with slowly scaling complexity with the number of vectors-combinations, 
since each updated vector weight is just compared against the weight of the current head,  
%combination 
 and only takes its place if it is higher. % or else it is placed under it. 
Retrieved packets are subsequently combined using the NCRAWL \emph{encode} 
library call, and the resulting encoded packet is scheduled for transmission.
 
{\bf Implementation considerations for algorithm 2:} 
 This algorithm is similar to algorithm 1, however it involves an additional 
acknowledgment scheme logic. 
Therefore, for each flow NCRAWL now maintains two queues: 
 (a) one with new incoming packets, and 
 (b) one with packets that have been successfully 
logged as keys by fellow nodes, but have not reached their ultimate 
destinations\footnote{For example this could be due to the fact that the destination failed to decode properly a previously sent encoded packet.}. 
Algorithm  2  exploits the 
NCRAWL acknowledgment scheme facility; this process groups the packet 
acknowledgment tokens, which have been created for 
outgoing packets combined together in the same encoded packet. 
This information is provided by NCRAWL to the   
developer. Algorithm 2  directly sends packets that have not 
yet reached their destinations; %, and have not been acknowledged 
%as %good 
%keys either; 
  such packets are not reconsidered for encoding. 
 % without reconsidering them for encoding. 
 However, the algorithm considers favorable queues and ``unknown" queues for the same flow separately, when forming vectors. 
 Note that the vectors formed with this algorithm scale intrusively, 
compared to the simple maxweight algorithm described previously. Throughout our measurements we only consider the scenario of two flows and thus avoid the arising complexity. This issue is expected to be resolved in the future using the NCRAWL framework.

{\bf Algorithm 3 in NCRAWL:} 
 For the implementation of the third algorithm we simply need to create vectors, (i.e. controls or queue combinations) for which the decoding
probability is nonzero, according to the user-defined threshold $\delta$ and the channel quality. 
As soon as packets are 
available in all %flow 
queues that constitute a vector, they are combined
and transmitted at once, without considering or updating the queue backlogs. This
algorithm %tries to achieve the maximum coding gain and thus, it will choose 
selects controls that mix the largest possible number of packets each time.

 We should note here that NCRAWL does not use any time-threshold policy 
 towards increasing the backlog size of the incoming packet queues, before 
 % arriving packet queue backlogs before 
 deciding to send outgoing packets. On the other hand, COPE adopts such a design decision. 
 % like it happens with {COPE}. 
 With NCRAWL, %By default
queue backlogs will increase when the relay's outgoing packet rate is smaller than the incoming packet rate. In such cases, NC proves to be a panacea for the router stability; if the NC algorithmic operations are supported by a lightweight implementation, the router capacity can be truly increased, as our measurements suggest. 
%%%%%%%%%%%%%%%%%%%%%%%%%%%%%%%%%%%%%%%%%%%%%%%%%%%%%%%%%%
%%%%%%%%%%%%%%%%%%%%%%%%%%%%%%%%%%%%%%%%%%%%%%%%%%%%%%%%%%
%%%%%%%%%%%%%%%%%%%%%%%%%%%%%%%%%%%%%%%%%%%%%%%%%%%%%%%%%%
\vspace{-0.1in}
\section{Evaluating our Framework}
\label{sec:measurements}

In this section, we %assess the performance of 
 evaluate 
NCRAWL %framework 
in conjunction with scheduling algorithms (NCRAWL + alg1, NCRAWL + alg2 and $\delta$--FTP) described in section \ref{sec:case},  
in terms of both throughput and resource utilization. We begin by describing 
the wireless testbed infrastructure\footnote{Our motivating experiments discussed in section \ref{sec:introduction} were performed on 2 different testbeds. We have evaluated NCRAWL on both testbeds; here we present results for one of them.} 
and the configurations that we 
used to deploy experiments. Next, we quantify the 
{CPU} overheard that is introduced by each NCRAWL 
processing stage, under maximum traffic loads, and we compare 
total {CPU} utilization to: (i) the public COPE implementation 
that uses an explicit acknowledgment scheme and (ii) legacy IEEE 802.11b-g.
Following, we demonstrate that NCRAWL can support theoretical 
gains even when coding opportunities lead to more than 2-packet 
combinations. Finally, we deploy    experiments that 
demonstrate how the proposed algorithms perform in cases with variable link qualities and different rates. % in the neighborhood of the relay.

\subsection{Experimental setup}

Our testbed is comprised of 20 ORBIT-like nodes, deployed both indoors and outdoors. 
Each node consists  of one 1{GHz} 386 processor, 512MB of {RAM}, two ethernet ports and two miniPCI slots which are used to host two  AR5212 Atheros 802.11a/b/g WiFi 
cards. % that support IEEE 802.11a, b and g modes.
All the nodes are connected through wired Ethernet with the testbed's 
server -  console. On console, we have all the required testbed 
services running as well as the NCRAWL deployment scripts that 
we described in section~\ref{subsec:deploy_NCRAWL_experiments}. 
 For conducting  throughput measurements we use the {iperf} bandwidth meter tool, \cite{iperf}. 
For {CPU} occupancy measurements  we appropriately instrument  
NCRAWL %framework  
 with the Linux \emph{getrusage} system call, which  
accurately estimates {CPU} usage time. We   place  several {getrusage} 
calls at the borders of each processing stage, we record  the average usage time 
of each stage and we compare  it to the whole NCRAWL system usage time. 
 We have repeatedly performed all of our experiments late at night, in order to avoid interference from collocated networks.

\subsection{CPU occupancy measurements}

In order to measure the efficiency of our framework in terms of CPU occupancy, we compare it to the case of running COPE \cite{cope}, as well as %only 
the legacy IEEE 802.11 protocol.

\begin{figure*}[h]
\begin{center} \hspace{-0.12in}
\parbox{1.5in} {
%\vspace{0.06in}
    \centerline{\subfigure[%CPU occupancy in b mode
    ]{\includegraphics[scale=.19]{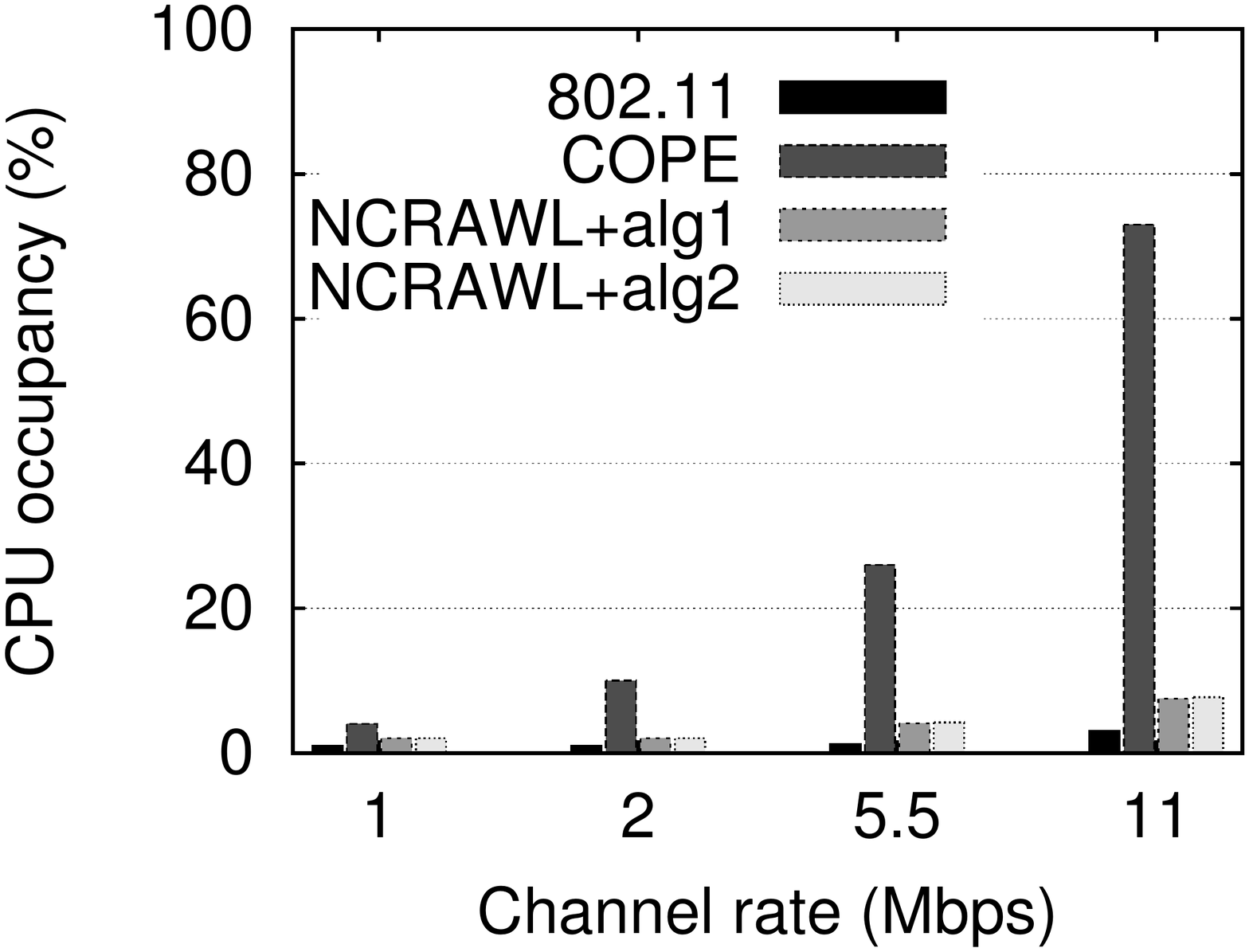}}}
 %   \vspace{-0.1in}
 %   \caption{(a)}
 %   \label{fig:CPU_b}
}
\makebox[.15in] {}
\parbox{1.5in} {
% \vspace{0.06in}
    \centerline{\subfigure[%CPU occupancy in g mode
    ]{\includegraphics[scale=.19]{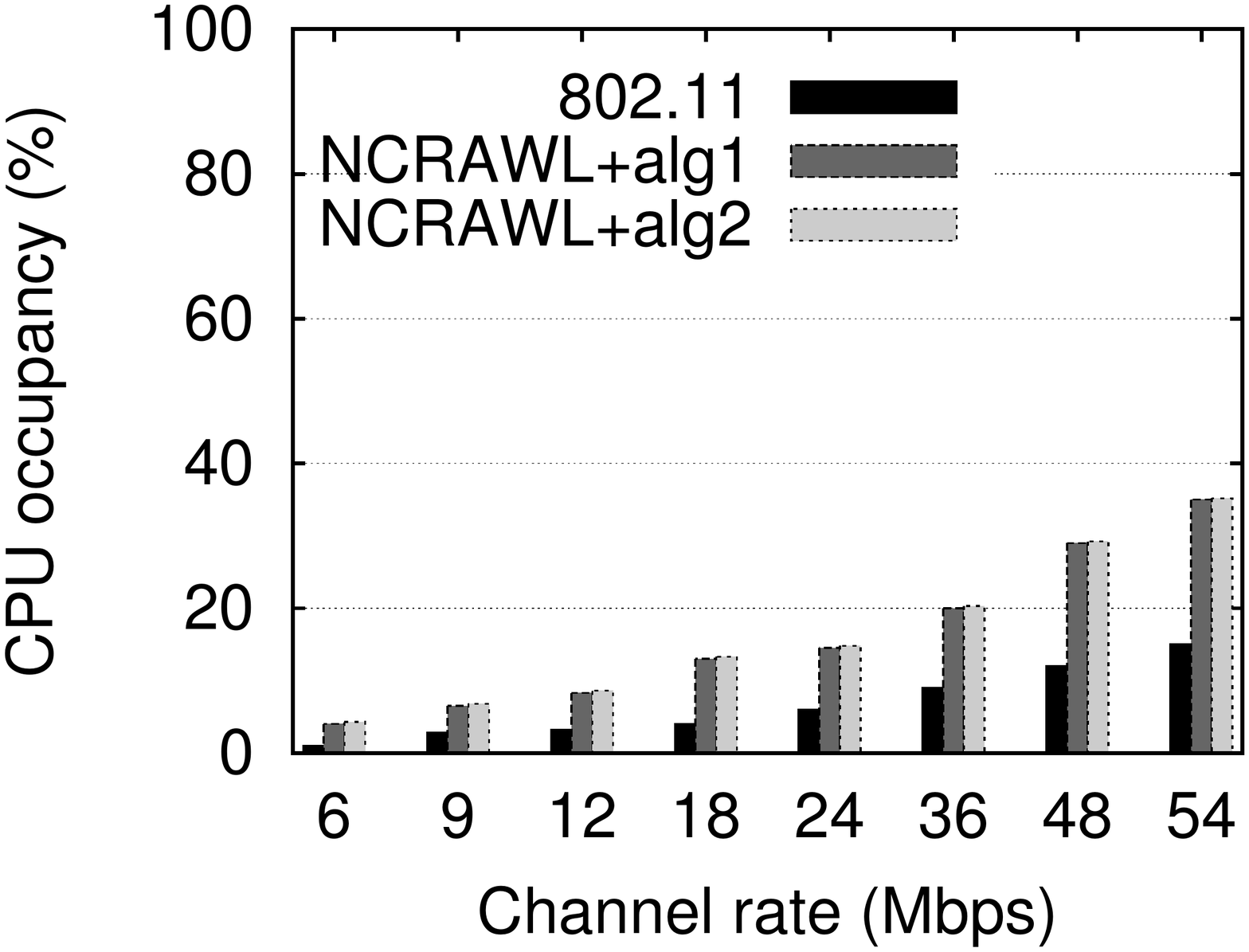}}}
   % \vspace{-0.1in}
 %   \caption{(b)}
 %   \label{fig:CPU_g}
}
\makebox[.15in] {}
\parbox{1.8in} {
\vspace{0.1in}
    \centerline{\subfigure[%Breakdown of CPU occupancy
    ]{\includegraphics[scale=.17]{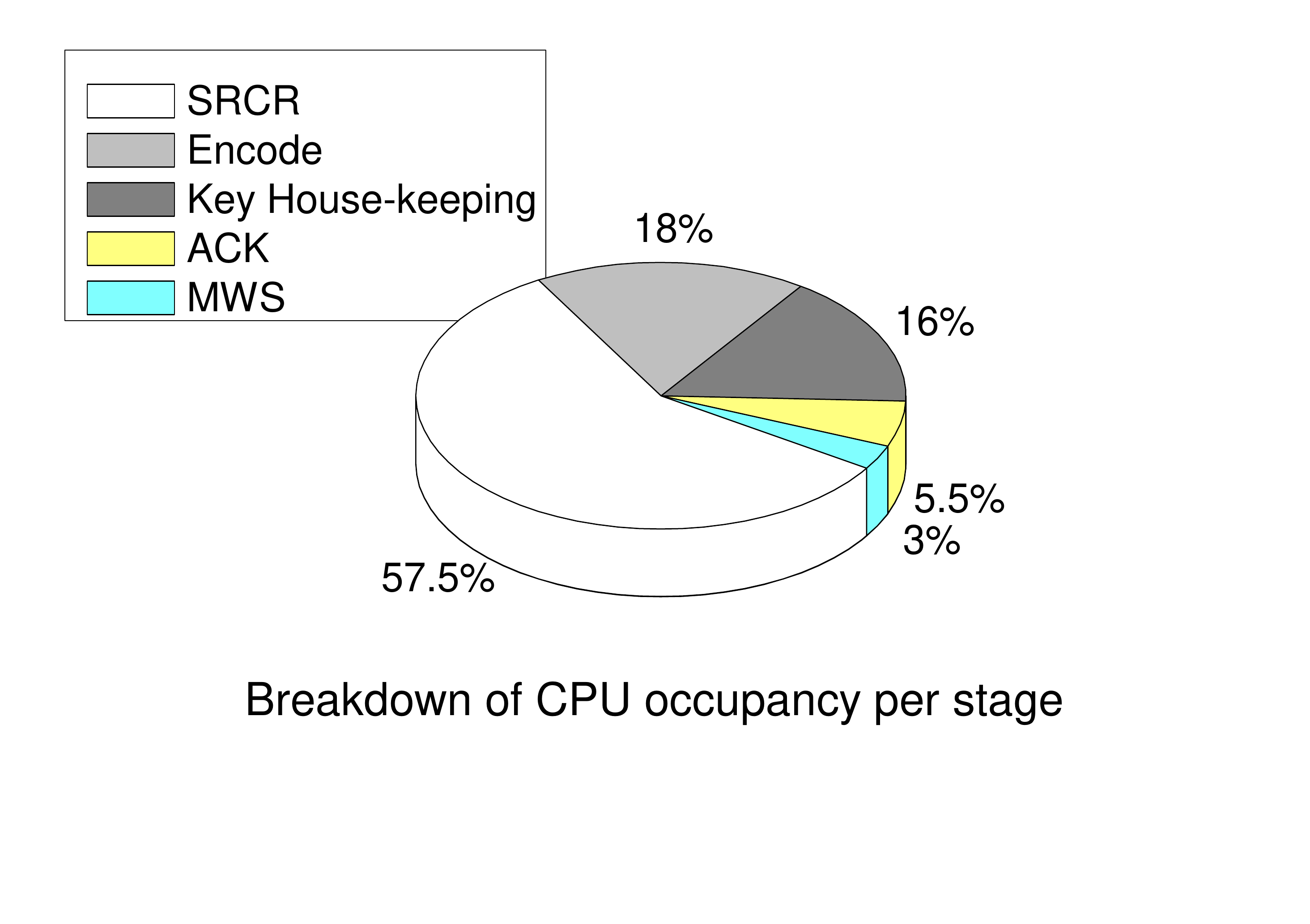}}}
    %\vspace{-0.1in}
 %   \caption{(c)}
  %  \label{fig:CPU_pie}
}
\makebox[.15in] {}
\parbox{1.2in} {
%\vspace{0.1in}
\hspace{-0.2in}
    \centerline{\subfigure[%Breakdown of CPU occupancy
    ]{\includegraphics[scale=.19]{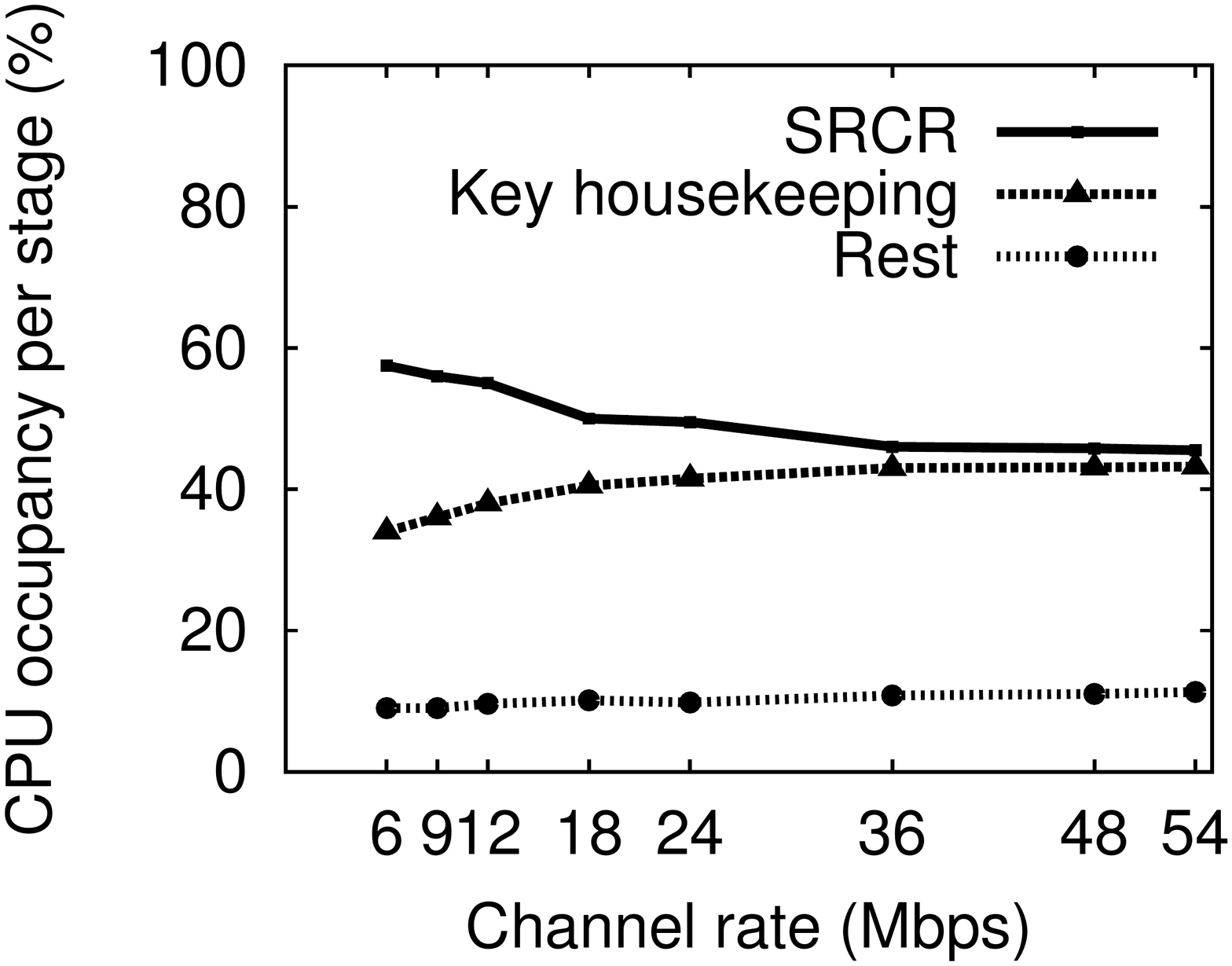}}}
    %\vspace{-0.1in}
 %   \caption{(c)}
  %  \label{fig:CPU_pie}
}
%\caption{CPU occupancy for several cases}
%\label{fig:CPU}
\end{center}
 %\vspace{-0.125in}
\end{figure*}

%%%%%%%%%%%%%%%%%%%%%%%%%%%%%%%%%%%%%%%%%%%%%%%%%%%%%%%%%%%%%%%%%%%%%%%%%%%%%%%%%%%%%%%%%%%%%%%%%%%%%%%%%%%%%%%%%

\begin{figure*}[h] \vspace{-0.12in}
\begin{center} \hspace{-0.12in}
\parbox{1.3in} {
%\vspace{0.06in}
    \centerline{\subfigure[%Breakdown of CPU occupancy vs rate
    ]{ \includegraphics[scale=.19]{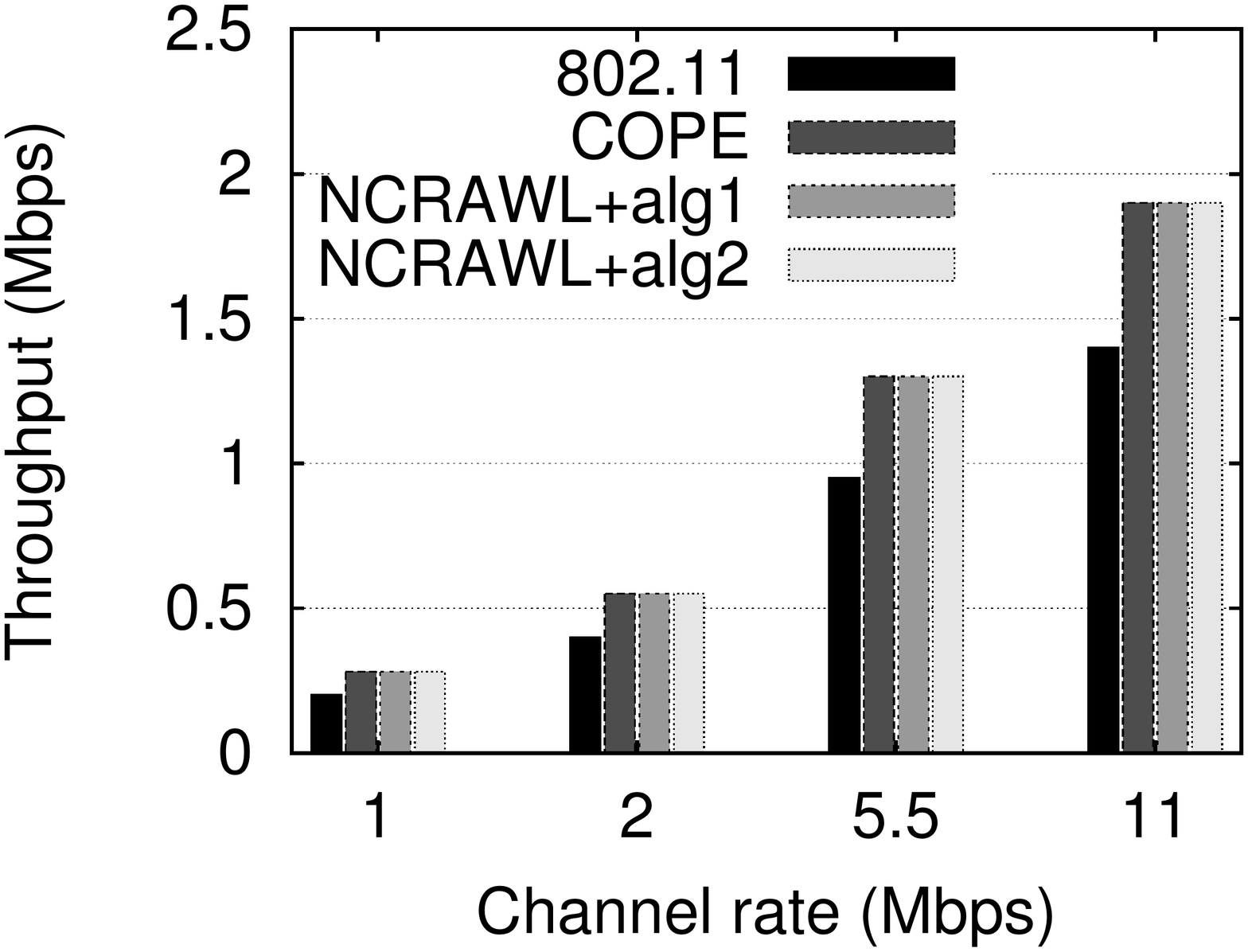}}}
 %   \vspace{-0.1in}
 %   \caption{(a)}
 %   \label{fig:CPU_b}
}
\makebox[.34in] {}
\parbox{1.3in} {
%\vspace{0.06in}
    \centerline{\subfigure[%Breakdown of CPU occupancy vs rate
    ]{ \includegraphics[scale=.19]{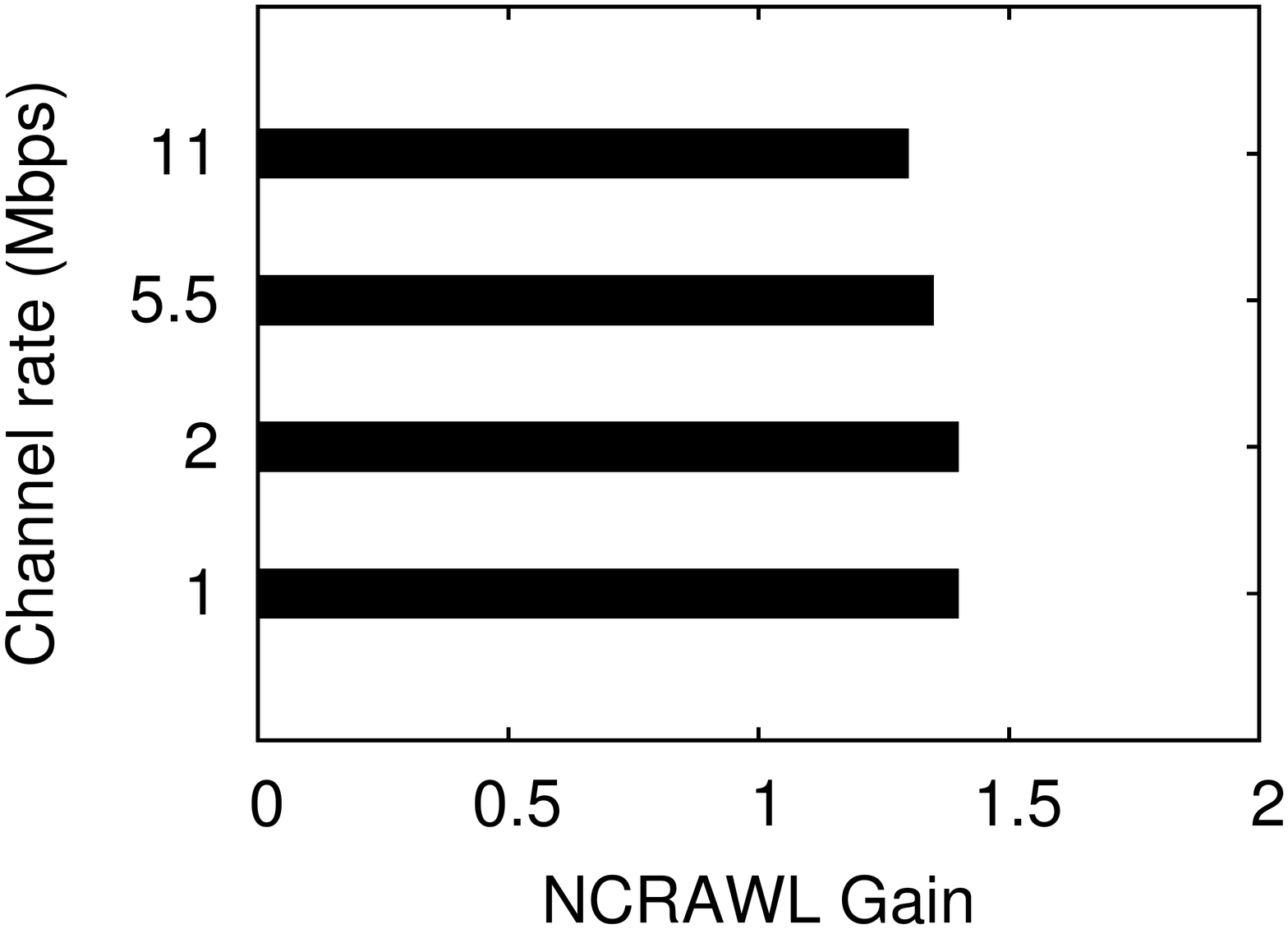}}}
 %   \vspace{-0.1in}
 %   \caption{(a)}
 %   \label{fig:CPU_b}
}
\makebox[.34in] {}
\parbox{1.3in} {
%\vspace{0.06in}
    \centerline{\subfigure[%Breakdown of CPU occupancy vs rate
    ]{ \includegraphics[scale=.19]{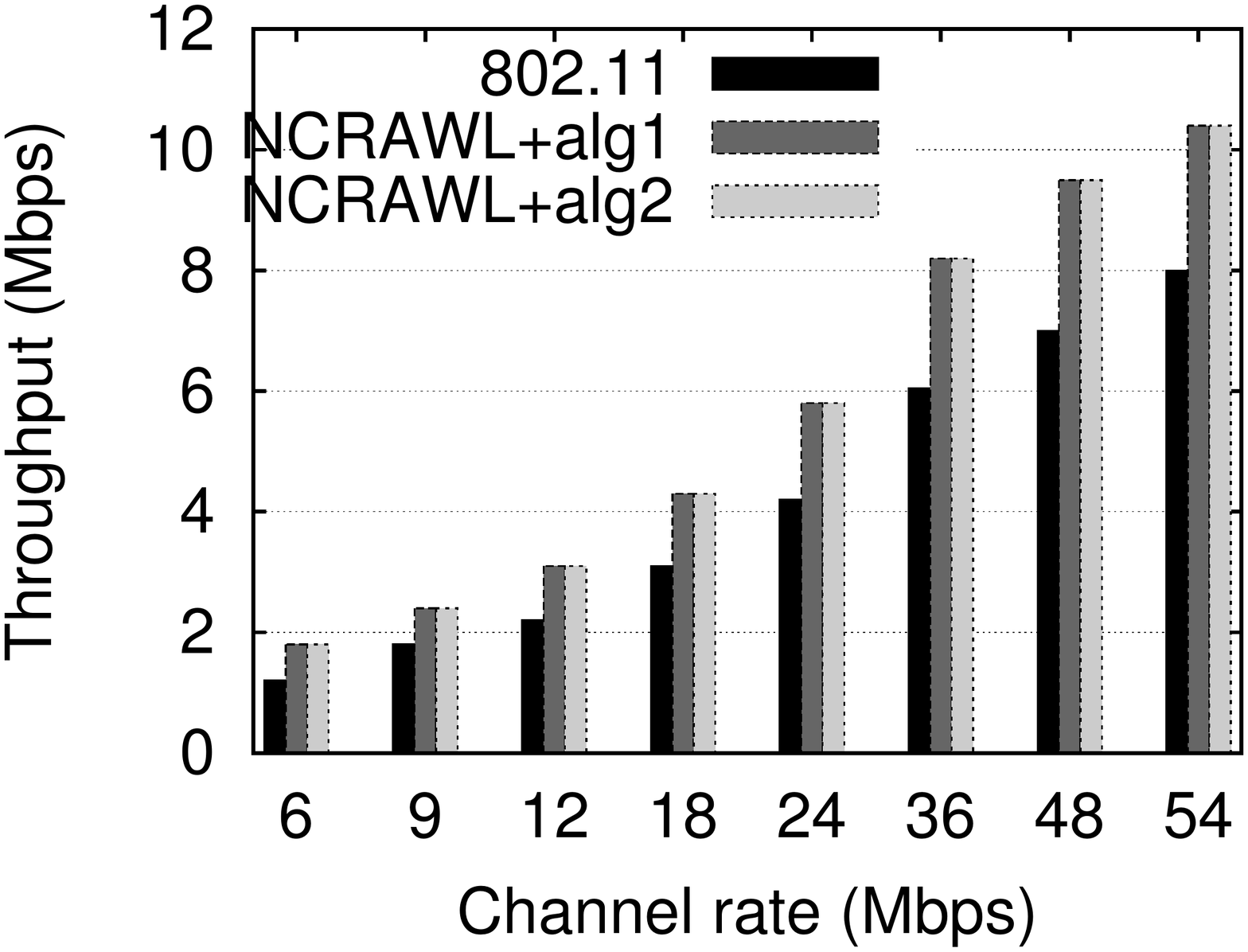}}}
 %   \vspace{-0.1in}
 %   \caption{(a)}
 %   \label{fig:CPU_b}
}
\makebox[.34in] {}
\parbox{1.3in} {
%\vspace{0.06in}
    \centerline{\subfigure[%Breakdown of CPU occupancy vs rate
    ]{ \includegraphics[scale=.19]{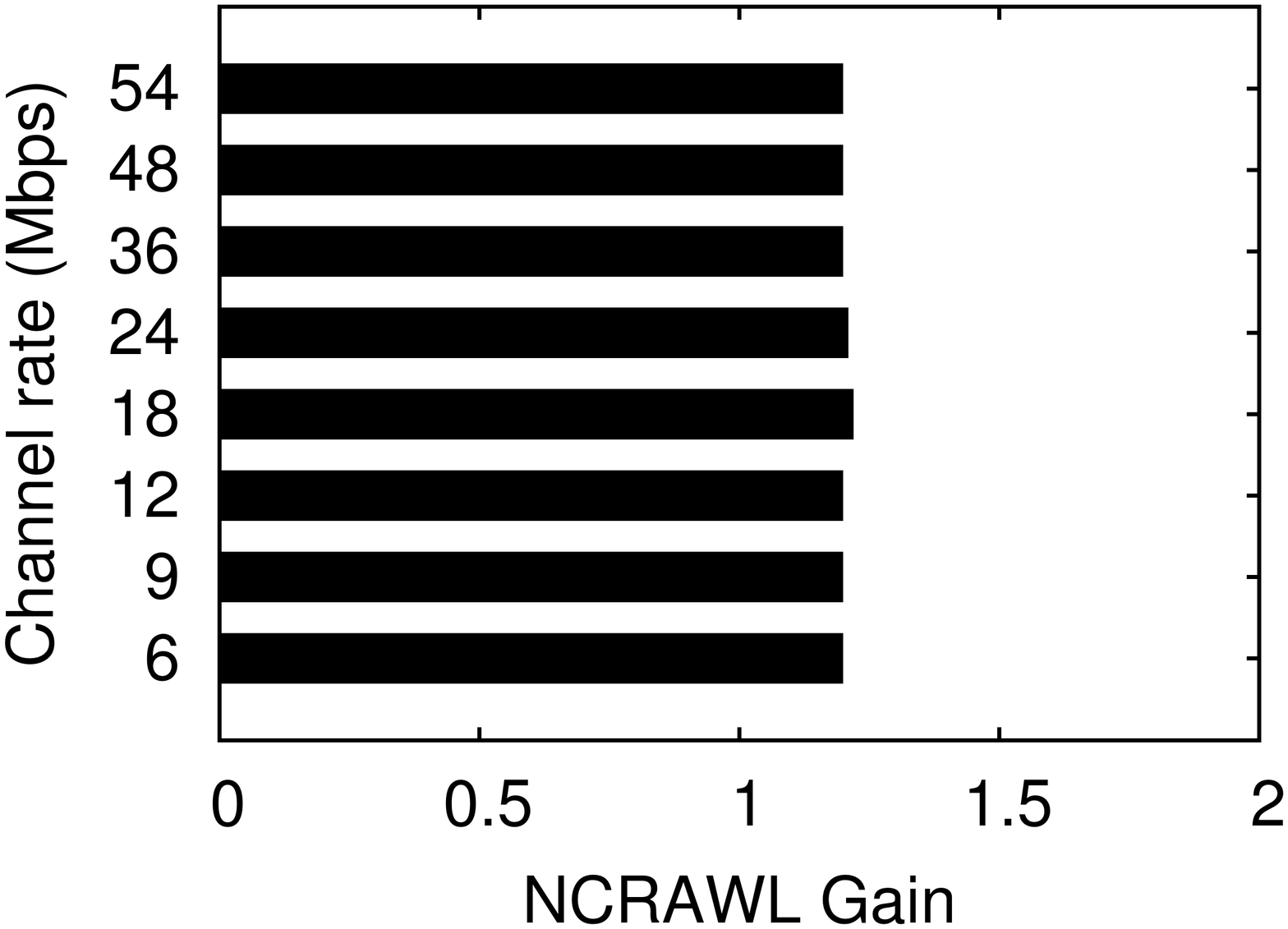}}}
 %   \vspace{-0.1in}
 %   \caption{(a)}
 %   \label{fig:CPU_b}
}
% \vspace{-0.2in}
\caption{Results in Alice-Relay-Bob topology.}
\label{fig:alice}
\end{center}
%\vspace{-0.15in}
\end{figure*}

%%%%%%%%%%%%%%%%%%%%%%%%%%%%%%%%%%%%%%%%%%%%%%%%%%%%%%%%%%%%%%%%%%%%%%%%%%%%%%%%%%%%%%%%%%%%%%%%%%%%%%%%%%%%%%%%%

\begin{figure*}[t!]
\begin{center} \hspace{-0.12in}
\parbox{1.3in} {
%\vspace{0.06in}
    \centerline{\subfigure[%Throughput vs \# flows
    ]{ \includegraphics[scale=.19]{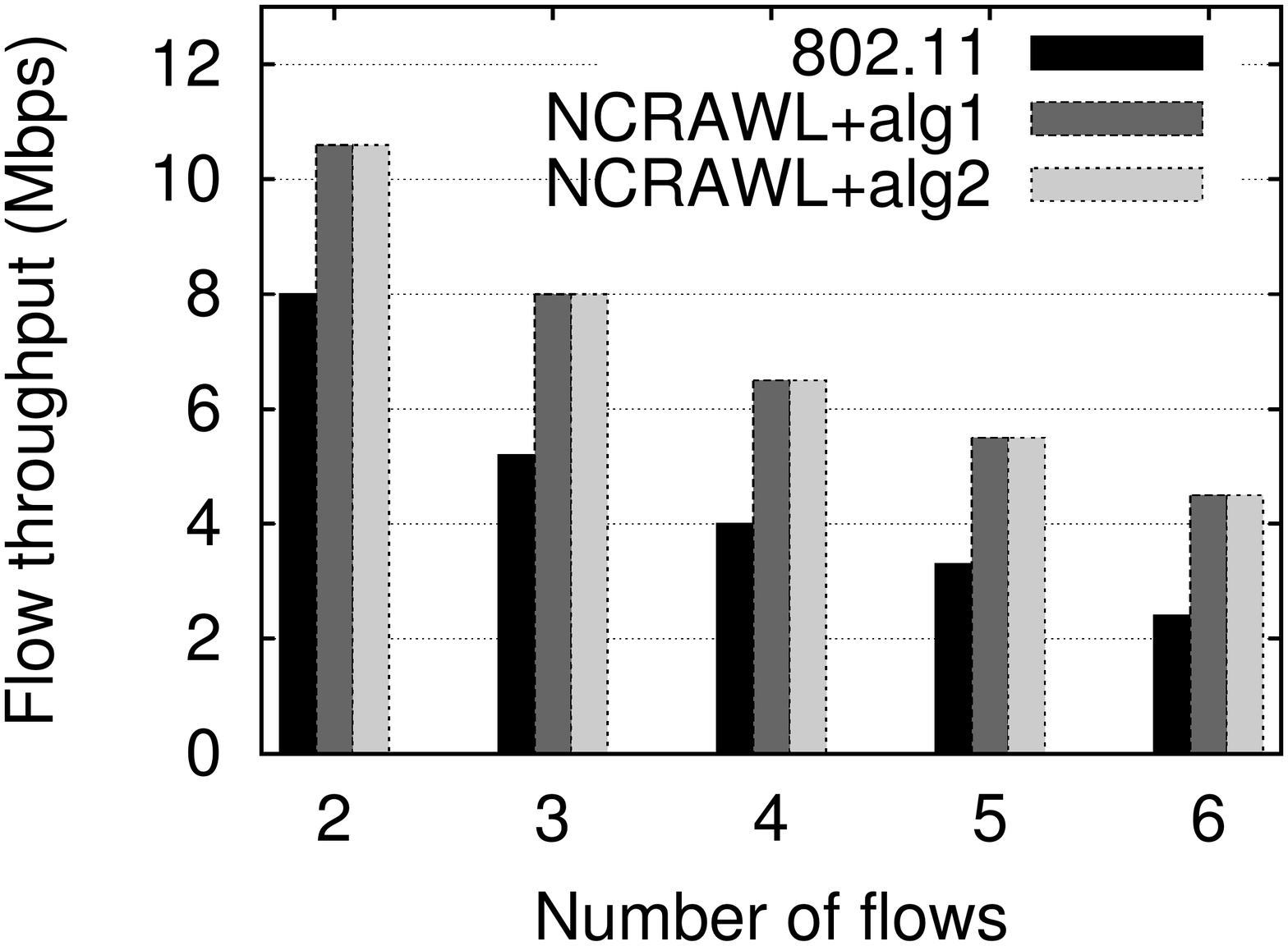}}}
 %   \vspace{-0.1in}
 %   \caption{(a)}
 %   \label{fig:CPU_b}
}
\makebox[.35in] {}
\parbox{1.3in} {
% \vspace{0.06in}
    \centerline{\subfigure[%Throughput vs q
    ]{\includegraphics[scale=.19]{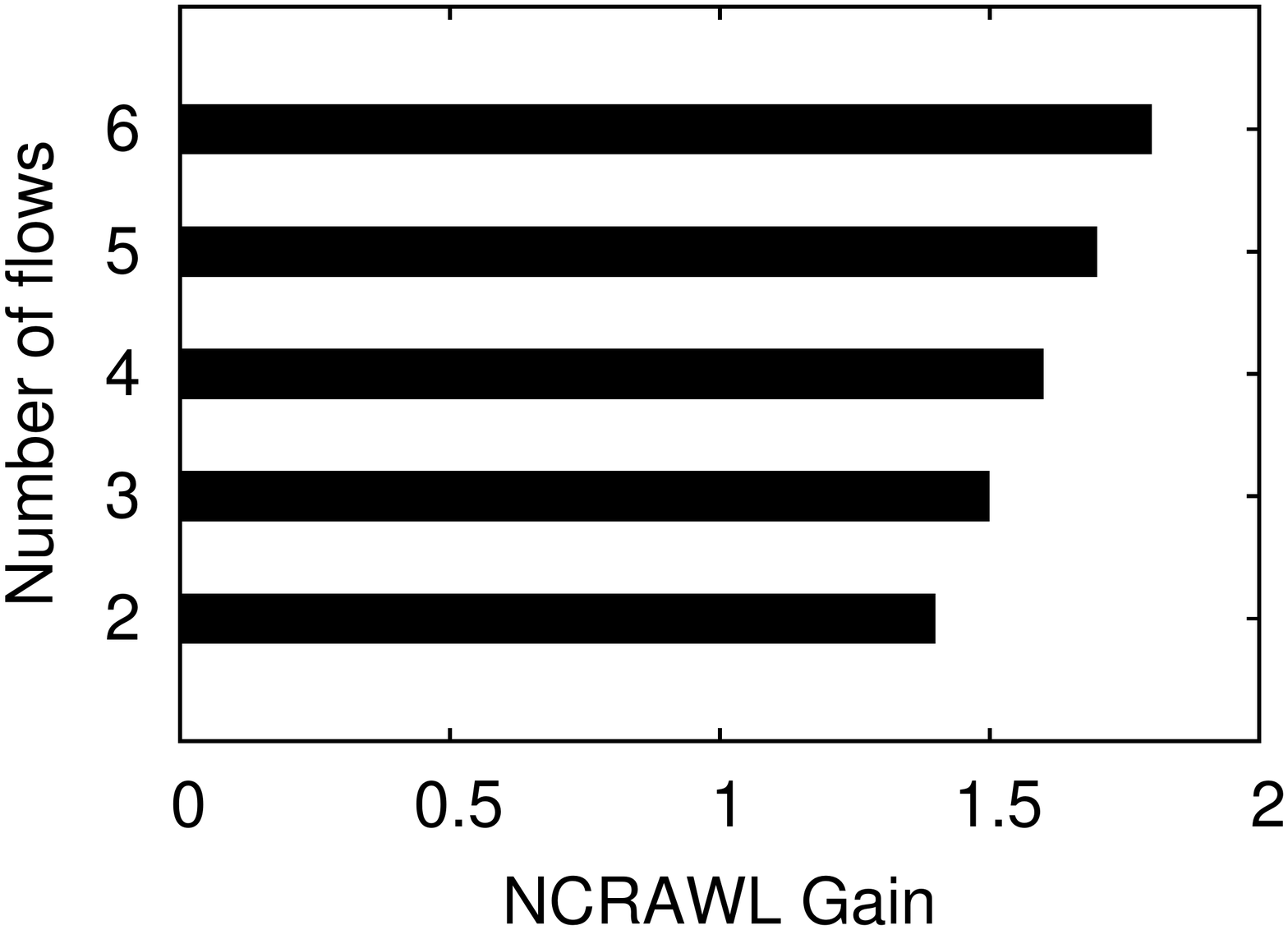}}}
%    \vspace{-0.1in}
 %   \caption{(b)}
 %   \label{fig:CPU_g}
}
\makebox[.35in] {}
\parbox{1.3in} {
% \vspace{0.06in}
    \centerline{\subfigure[%Throughput vs q
    ]{\includegraphics[scale=.19]{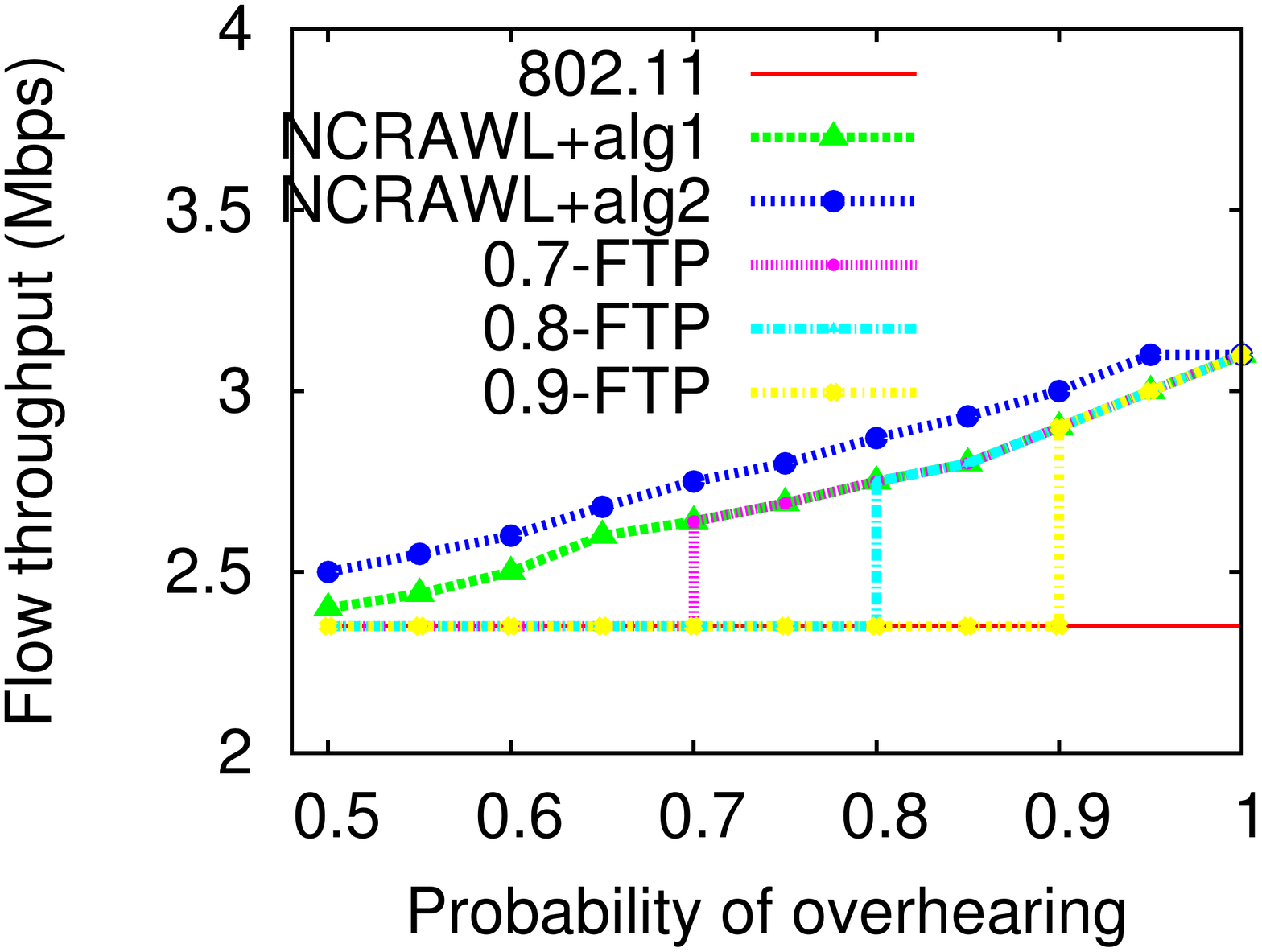}}}
%    \vspace{-0.1in}
 %   \caption{(b)}
 %   \label{fig:CPU_g}
}
\makebox[.35in] {}
\parbox{1.3in} {
%\vspace{0.06in}
    \centerline{\subfigure[%Throughput vs ch. rate
    ]{\includegraphics[scale=.19]{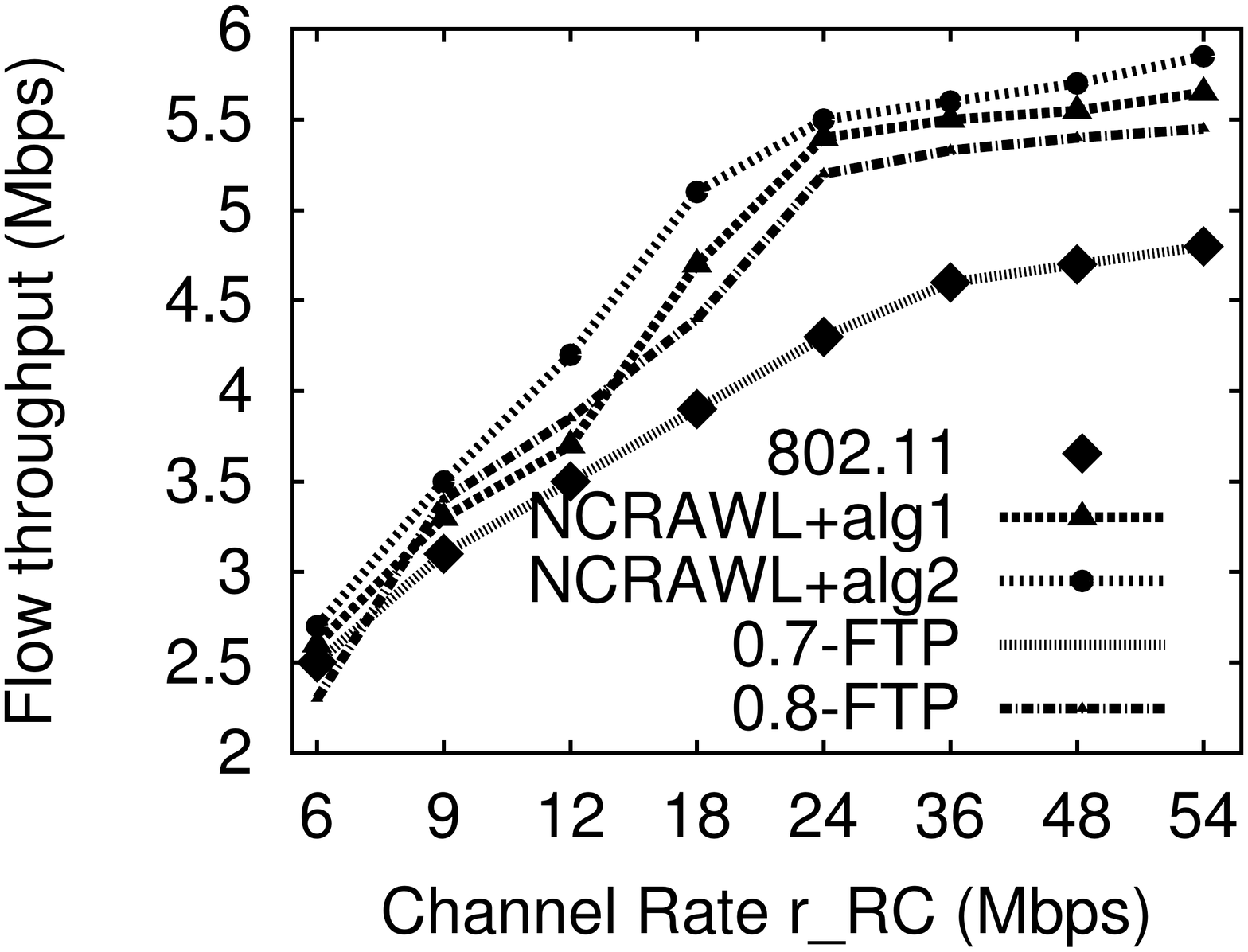}}}
%    \vspace{-0.1in}
  %  \caption{(c)}
  %  \label{fig:CPU_pie}
}
% \vspace{-0.2in}
\caption{Results in wheel topologies. } % (see the text for more information).}
\label{fig:wheel}
\end{center}
%\vspace{-0.13in}
\end{figure*}

{\bf NCRAWL is much more CPU friendly than COPE-based approaches:} 
%In particular, we
 We  invoke the Alice-Relay-Bob setting (see section \ref{sec:nc_scheme}) and we inject fully saturated traffic in both flows. 
We compare NCRAWL + alg1, NCRAWL + alg2, %with \ref{fig:alice}-a IEEE 802.11b and 
COPE and the plain 802.11, for the case of 802.11b;   figure \ref{fig:alice}-a depicts the results.   
% and with \ref{fig:alice}-b IEEE 802.11g. 
Note that COPE can support at most the IEEE 802.11b rate set as discussed in section \ref{sec:introduction}; for the sake of a fair comparison here, we use this mode of operation for NCRAWL as well.
% It is evident that our framework is lightweight and able to deliver throughput gains at high rates, which is a promising result.
We observe that NCRAWL makes use of the CPU resources in a very efficient manner: it reduces the CPU utilization by at least 2 and as much as 7 times compared to COPE  
 (we have validated these observations for the case of ER \cite{er} as well, which is based on COPE). 
Furthermore, we test NCRAWL for the case of 802.11g. Our measurements  (figure  \ref{fig:alice}-b) 
 suggest that NCRAWL does not need to occupy more than 37\% of the CPU resources for NC operations at 54 Mbps, with fully saturated UDP traffic! This implies that the design of NCRAWL  includes low additional overhead functions (as opposed to legacy 802.11). 

{\bf Evaluating  individual operations of NCRAWL:} 
Next, we deploy {getrusage} calls and measure the breakdown of CPU occupancy per processing stage (figure \ref{fig:alice}-c). 
The most CPU intensive operation is the SRCR stage (it contains legacy IEEE 802.11 operations as well). %; %which implies that 
 %NCRAWL carries a %implementation brings small extra burden to the whole process. In particular, 
 The most computationally heavy pieces of NCRAWL are the encode stage and the key house-keeping. Note  here  that these two lie at the heart of any NC system and in a way represent unavoidable costs. It should also be noted that the   processing stage of the scheduler remains at very low values and there is a certain percentage dedicated to dealing with ACKs. 
 Furthermore, as depicted in figure \ref{fig:alice}-d,   by increasing the channel rate (and thus the number of packets into the system per unit time), the coding stage increases in complexity disproportionally with the SRCR stage. This implies that the coding complexity increases faster than SRCR as the rate increases. 
 Nevertheless, %it seems that 
 for %very 
  high channel rates  the differences  are reduced. 
  This suggests that NCRAWL could potentially operate efficiently at much higher channel rates, such as with 802.11n systems. We plan to test NCRAWL on MIMO networks in our future work. 
   
\subsection{Throughput measurements with UDP}
%In this subsection we show that our implementation supports theoretical gains in throughput. 
Next, we assess the ability of NCRAWL to approach the theoretically expected benefits of NC. 

{\bf Experiments with the simple Alice-Relay-Bob topology:}
%For this, 
 %we use the Alice-Relay-Bob topology and
 We calculate and measure the maximum throughput for both symmetric   flows, such that the system remains stable (i.e. the queues do not rise more than a large permissible number). 
Figures \ref{fig:alice}-e, \ref{fig:alice}-f, \ref{fig:alice}-g and \ref{fig:alice}-h show the results. 
 %The comparison is again similar to the previous subsection. 
 Note that since the receivers   always have the proper keys (these are the keys from their own transmitted packets \cite{proutiere}), decoding is always possible and thus algorithm 1, algorithm 2 and COPE are optimal in this setting. 
 In each case, a gain in throughput of $\frac{4}{3}$ is identified, which matches the theoretical for this topology. 
 Our measurements suggest that 
 %with COPE at high channel rates suggest that the theoretical throughput is never achieved. 
%
COPE achieves the theoretical throughput for small rates, but %in our testbed 
 it fails to do so in higher rates. 
 Note that the public COPE code was initially available for 802.11b only; 
 while we carefully %tried to 
  modified COPE to operate at 802.11g rates, 
  we observed that such modifications lead to a very unstable system when rates higher than 18 Mbps are used. A closer look at certain individual components of the COPE implementation revealed that the reason for this instability is the  excessive overhead induced by the NC system operations (as discussed earlier). 
% alterations were necessary for these measurements. 
For this reason we do not explicitly
compare COPE here at these high rates. Nevertheless, from these measurements  
 one can realize %with a simple linear digression 
 that COPE cannot provide benefits at 
 rates higher than 18 Mbps, due to the tremendous CPU processing overheads that its design  incurs. 
 In contrast, NCRAWL manages to %actually 
 reach the theoretical gain at high channel rates (e.g. at 54 Mbps), as shown in figures \ref{fig:alice}-f and \ref{fig:alice}-h. % is then an important achievement of our framework.

{\bf The case for wheel topologies:} 
Furthermore, we scale the number of flows (see figures \ref{fig:wheel}-a and \ref{fig:wheel}-b); the topology is an $\frac{x}{2}$--wheel. % (see section \ref{sec:nc_scheme}). 
The theoretical gain in this case is $\frac{2x}{x+1}$ where $x$ is the number of flows combined at the downlink. Our measurements support the theoretically predicted gain at the channel rate of 54 Mbps. We observe the per flow throughput naturally drops, as the number of flows increases, but the aggregate throughput increases. The gain (figure \ref{fig:wheel}-b)) is an increasing function of flows and approaches asymptotically 2; note that this is perfectly aligned   with the findings in \cite{proutiere} as well.  
 % It is important to stress 
  Note also that in $\frac{x}{2}$--wheel topologies, piggybacking is not available since there is no return flow from the receivers. NCRAWL is able to select the appropriate ACKing method and the results show that the overhead incurred is negligible.

{\bf Experiments with cross topologies:} 
We now present two more cases of interest that can appear in realistic environments. 
We setup % the experiments on 
 various {\em cross} topologies with nodes in different locations across our testbed; we activate the flows Alice-Relay-Chloe and Bob-Relay-David. The arrivals are again chosen in a symmetric way, i.e. the arrival rate of the one flow is equal to the other.

$\bullet$ In the first case (figure \ref{fig:wheel}-c),  David overhears Alice's uplink transmissions with probability 1 and Chloe hears Bob with probability $q$. The rates of all links are equally set to 12Mbps (the channel rate is not important in this experiment). We measure the highest throughput that guarantees queue stability while varying the probability $q$, by considering different node locations. 
We compare NCRAWL+alg1, NCRAWL+alg2 and IEEE 802.11g as well as $\delta$--FTP for $\delta=\{0.7,0.8,0.9\}$ (see section \ref{sec:case} for description). The results demonstrate the superiority of NCRAWL+alg2, which is able to deliver the maximum throughput in each case. Evidently, our framework in combination with the proposed scheduling algorithms is able to effectively handle %in an effective way 
the several link quality conditions. 

$\bullet$  In the second case (figure \ref{fig:wheel}-d), %we use the same topology. This time 
 the overhearing probability from Bob to Chloe is set to $q=0.7$. All channel rates are set to 24Mbps with the exception of the link Relay-Chloe which is varied. Our measurements demonstrate the inefficiency of policies oblivious to rates like the $\delta$--FTP. In this case, the choice of a small value for $\delta$ is penalized when the Relay-Chloe link is slow enough. Instead NCRAWL+alg2 is able to handle in an effective way the several rate and link conditions and deliver important throughput gains. 
From figures \ref{fig:wheel}-c and \ref{fig:wheel}-d  
we also observe that 
 %one can also compare NCRAWL+alg1 and NCRAWL+alg2. Once
 given that  
  overhearing links are not perfect in terms of PDR, NCRAWL+alg2 always outperforms NCRAWL+alg1, since it is able to use feedback information.

\subsection{Performance with TCP traffic}

Finally, we assess the efficacy of NCRAWL in scenarios with TCP traffic. 
%So far, all the experiments were carried through using iperf (an application layer tool) over UDP protocol. 
In \cite{cope}, experiments with TCP have demonstrated a loss in efficiency due to packet losses and reordering. 
 %We also deployed TCP experiments. Initially, using the 
First, throughout our experiments with the Alice-Relay-Bob topology, where no losses or delays are incurred, the throughput is reduced due to the additional TCP  overheads. We observe that when the 54 Mbps rate is used, the per flow throughput rate is 7 Mbps for plain 802.11 and 8.5 Mbps for NCRAWL+alg1. 
 A slight loss in NC gain is observed; this   is the result of mixing TCP ACKs with data packets. The same gain is obtained for all the other available bit rates. % tested. 

Furthermore, 
 %In a second experiment, we used 
 we perform experiments with {\em half-cross} topologies, where flows are unidirectional (from Alice to Chloe and from Bob to Dave),   
 with probabilities of overhearing $q_{AD}=q_{BC}=0.7$ and several channel rates. In this case, NCRAWL+alg1 achieves a slightly lower throughput than IEEE 802.11. This is due to the fact that some packets are not correctly decoded at the destination and therefore they arrive delayed and out of order. This causes abrupt reactions from TCP and leads to throughput reduction. When adding the reordering module of COPE \cite{cope}, the packets arrive always in order, however this module increases the delay for each packet. This in turn is interpreted by TCP as congestion; it ends up in TCP window increments, and thereby decreases performance. 
NCRAWL is not optimized to cooperate with TCP at this point and thus, it faces the common problems of TCP in wireless networks. Improving this component is the main goal of our future work.
%%%%%%%%%%%%%%%%%%%%%%%%%%%%%%%%%%%%%%%%%%%%%%%%%%%%%%%%%%
%%%%%%%%%%%%%%%%%%%%%%%%%%%%%%%%%%%%%%%%%%%%%%%%%%%%%%%%%%
%%%%%%%%%%%%%%%%%%%%%%%%%%%%%%%%%%%%%%%%%%%%%%%%%%%%%%%%%%
\vspace{-0.1in}
\section{Conclusions}
\label{sec:conclusions}

%In this paper, we 
 We design and develop  NCRAWL. Our  framework 
%a Network Coding framework for Rate Adaptive Wireless Links. 
%
%NCRAWL 
 is an extended, generic NC framework that can be used to quickly develop networking systems in order to evaluate intersession NC and/or scheduling  algorithms,   entirely based on the implicit (probabilistic) acknowledgment that a packet can get decoded at 
the destination. The design of NCRAWL involves all the common processing steps that are always 
needed to implement such algorithms; these steps  have been abstracted such 
that designers %developers 
need to simply focus only on the  implementation of their algorithms. 
Our measurements %on a wireless testbed 
demonstrate that NCRAWL is a powerful NC development system. 
It offers significant throughput benefits  %router throughput performance improvements of the order of 50\%,
% as compared to previously proposed NC frameworks. 
even at high channel rates.  %in cases of channel rates as high as 54Mbps.

% {\bf Acknowledgments:} The work of D. Syrivelis and L. Tassiulas was supported by the European Commision project STREP-FP7-INFSO-ICT-215252:N-CRAVE. The work of G. S. Paschos and L. Georgiadis was supported by the project STREP-FP7-INFO-ICT-224218: OPNEX.

\end{document}